\documentclass[longauth]{aa}
\usepackage{graphicx}
\usepackage{txfonts}
\usepackage{lipsum}
\usepackage{subcaption}         
\usepackage{lscape}             
\usepackage{placeins}           
\usepackage{booktabs}

\usepackage[
  colorlinks=true,
  linkcolor=blue,
  citecolor=blue,
  urlcolor=blue
]{hyperref}

\newcommand{\TV}{Department of Physics, University of Rome Tor Vergata, via della Ricerca Scientifica 1, 00133, Rome, Italy}
\newcommand{\SAP}{Department of Physics, University of Rome La Sapienza, P.le Aldo Moro 2, 00185 Rome, Italy}
\newcommand{\PSU}{Department of Astronomy and Astrophysics, The Pennsylvania State University, 525 Davey Lab, University Park, PA 16802, USA}
\newcommand{\CMU}{McWilliams Center for Cosmology and Astrophysics, Department of Physics, Carnegie Mellon University, Pittsburgh, PA 15213, USA}
\newcommand{\UNAM}{Instituto de Astronom{\'\i}a, Universidad Nacional Aut\'onoma de M\'exico, Apartado Postal 70-264, 04510 M\'exico, CDMX, Mexico}
\newcommand{\UNAMENS}{Instituto de Astronom{\'\i}a, Universidad Nacional Aut\'onoma de M\'exico, km 107 Carretera Tijuana-Ensenada, 22860 Ensenada, Baja California, México}
\newcommand{\ASU}{School of Earth and Space Exploration, Arizona State University, Tempe, AZ 85287, USA}
\newcommand{\IOFFE}{Ioffe Institute, 26 Politekhnicheskaya, St.Petersburg, 194021, Russia}
\newcommand{\CAG}{Department of Physics, University of Cagliari, SP Monserrato-Sestu, km~0.7, 09042 Monserrato, Italy}
\newcommand{\IAA}{Instituto de Astrof\'isica de Andaluc\'ia (IAA-CSIC), Glorieta de la Astronom\'ia s/n, 18008 Granada, Spain}
\newcommand{\MAL}{Ingeniería de Sistemas y Automática, Universidad de Málaga, Unidad Asociada al CSIC por el IAA, Escuela de Ingenierías}
\newcommand{\SSD}{Space Science Division, Naval Research Laboratory, Washington, DC 20375, USA}
\newcommand{\DOA}{Department of Astronomy, School of Physics,
Peking University, Beijing 100871, China}
\newcommand{\KIAA}{Kavli Institute for Astronomy and Astrophysics, Peking University, Beijing 100871, China}
\newcommand{\MUS}{School of Physics and Astronomy, Monash University, Clayton, VIC 3800, Australia}
\newcommand{\MUC}{OzGrav: The ARC Centre of Excellence for Gravitational Wave Discovery, Australia}
\newcommand{\NVNC}{Nevada Center for Astrophysics, University of Nevada, Las Vegas, NV 89154, USA}
\newcommand{\NNDoPA}{Department of Physics and Astronomy, University of Nevada, Las Vegas, NV 89154, USA}
\newcommand{\NAOC}{National Astronomical Observatories,
Chinese Academy of Sciences, Beijing 100012, China}
\newcommand{\UH}{Faculty of Science, University of Helsinki, Gustaf  Hallströmin katu 2, FI-00014 Helsinki, Finland}
\newcommand{\IPT}{Institute of Physics and Technology, Ural Federal  University, Mira str. 19, 620002 Ekaterinburg}
\newcommand{\ISA}{Ingeniería de Sistemas y Autom\'atica, Universidad de M\'alaga, Unidad Asociada al CSIC por el IAA, Escuela de Ingenier\'ias Industriales, Arquitecto Francisco Pe\~nalosa, 6, Campanillas, 29071 M\'alaga, Spain}
\newcommand{\SASS}{School of Astronomy and Space Sciences, University of Chinese Academy of Sciences, Beijing 100049, People’s Republic of China}
\newcommand{\ALA}{Los Alamos National Laboratory, Los Alamos, NM 87545, USA}
\newcommand{\HKI}{The Hong Kong Institute for Astronomy and Astrophysics, University of Hong Kong, Pokfulam Road, Hong Kong, China}
\newcommand{\HKU}{Department of Physics, University of Hong Kong, Pokfulam Road, Hong Kong, China}
                                
\begin{document}

   \title{Exploring the connection between compact object mergers and fast X-ray transients}

   \subtitle{The cases of LXT~240402A and EP250207b}


   \author{R. L. Becerra\inst{1,2}
        \and Yu-Han Yang\inst{1}
        \and Eleonora Troja\inst{1}
        \and Massine El Kabir\inst{1,3}
        \and Simone Dichiara\inst{4}
        \and Niccolo Passaleva\inst{1,3}
        \and Brendan O'Connor\inst{5}
        \and Roberto Ricci\inst{1}
        \and Chris Fryer\inst{6}
        \and Lei Hu\inst{5}
        \and Qinyu Wu\inst{7,8}
        \and Muskan Yadav\inst{1}
        \and Alan M. Watson\inst{2}
        \and Anastasia Tsvetkova\inst{9,10}
        \and Camila Angulo-Valdez\inst{2}
        \and María D. Caballero-García\inst{11}
        \and Alberto J. Castro-Tirado\inst{11,12}
        \and C. C. Cheung\inst{13}
        \and Dmitry Frederiks\inst{10}
        \and Maria Gritsevich\inst{14,15}
        \and J. E. Grove\inst{13}
        \and M. Kerr\inst{13}
        \and William H. Lee\inst{2}
        \and Alexandra L. Lysenko\inst{10}
        \and Margarita Pereyra\inst{16}
        \and Anna Ridnaia\inst{10}
        \and Rubén Sánchez-Ramírez\inst{11}
        \and Hui Sun\inst{6}
        \and Dmitry Svinkin\inst{10}
        \and Mikhail Ulanov\inst{10}
        \and R. Woolf\inst{13}
        \and Bing Zhang\inst{23,24,25,26}
        }

   \institute{\TV
             \and \UNAM \and \SAP \and \PSU  \and \CMU \and \ALA \and \NAOC \and \SASS \and \CAG \and \IOFFE \and \IAA \and \ISA \and \SSD \and \UH \and \IPT \and \UNAMENS \and \ASU 
             \and \MAL  \and \DOA \and \KIAA \and \MUS \and \MUC \and \NVNC 
             \and \HKI \and \HKU \and \NNDoPA }

   \date{Received September 30, 2025}

\abstract
{The connection between compact object mergers and some extragalactic fast X-ray transients (FXRTs) has long been hypothesised but never ultimately established.}
{In this work, we investigate two FXRTs, the LEIA X-ray Transient LXT\,240402A and the Einstein Probe EP\,250207b, whose precise positions lie close to nearby ($z\!\lesssim\!0.1$) quiescent galaxies with a negligible probability of chance coincidence, identifying them as particularly promising cases of merger-driven explosions in the local Universe.}
{We used \textit{Chandra} to derive accurate localisations for both events and secure otherwise ambiguous associations with their optical counterparts. Deep optical and near-infrared observations with VLT, GTC, and LBT were performed to characterise the surrounding environment and search for kilonova emission, the hallmark of neutron star mergers. Complementary early-time X-ray monitoring with \textit{Swift} and \textit{Einstein Probe} was used to constrain the non-thermal afterglow.}
{We find that both FXRTs remain compatible with a compact binary merger progenitor, which produced low-mass ejecta and kilonova emission subdominant to the afterglow. However, alternative explanations such as a distant ($z\!\gtrsim\!1$) core-collapse supernova cannot be conclusively ruled out.}
{}


   \maketitle

\section{Introduction}
\label{ sec:intro }

Mergers of two compact objects, either two neutron stars (NSs) or an NS and a black hole (BH), are widely recognised as the progenitors of most short-duration gamma-ray bursts (GRBs; \citealt{Eichler1989, Gehrels2005, Lee2007, Berger2014}). 
The multi-messenger detection of the NS merger GW170817, together with GRB~170817A and the optical/infrared kilonova (AT2017gfo), cemented this long-standing hypothesis by directly linking a NS merger to prompt high-energy emission and r-process nucleosynthesis \citep{Abbott2017PRL,Abbott2017MMA}. 
In this framework, a narrowly collimated relativistic jet launched by the merger remnant produces the initial gamma-ray phase \citep{Piran2004, Kumar2015}, whereas the sub-relativistic neutron-rich ejecta unbound during or after the merger power the kilonova via radioactive heating \citep{LiPaczynski1998,BarnesKasen2013,Tanaka2013,MetzgerLRR}.

More recently, mounting observational evidence has indicated that compact-object mergers are also likely progenitors of some long duration GRBs
\citep[e.g.][]{Gehrels2006, Zhang2007, Troja2022, Rastinejad2022, Yang2024,Eappachen2025}, demonstrating that the merger remnant, either an accreting BH or a massive NS, can produce classical short-duration($\lesssim$2 s) emission and also power minute-long gamma-ray transients
The unusual phenomenology of these hybrid GRBs has in turn motivated consideration of  alternative compact binary progenitor channels - most notably NS - white dwarf (WD) mergers - and the formation of a massive proto-NS to account for the prolonged high-energy emission \citep{King2007,Yang2022,Sun2025}.
Besides progenitor types, several other factors, such as central engine, jet geometry, and emitter physics,  contribute to the observed GRB duration, further limiting the reliability of duration-based classification \citep{Zhang2025}. 

Independent evidence of sustained central-engine activity in merger-driven outbursts comes from the phenomenology of short GRBs and their X-ray afterglows. A subset of short GRBs exhibits a temporally extended ($\sim$100\,s) and spectrally softer ($\lesssim$50 keV) emission component following the initial spike \citep{NorrisBonnell2006, Gehrels2006, Sakamoto2011, Kaneko2015}. In the X-ray band,  plateau phases are observed over timescales of $\sim10^2$–$10^3$~s \citep[e.g.][]{Fan2006,Rowlinson2013,Lu2015}. These features are often explained invoking long-lasting central engine activity well beyond the duration of the prompt gamma-ray phase. Proposed mechanisms include the spin-down of a highly magnetised neutron star (a 'magnetar' remnant) formed in the merger \citep[e.g.][]{Dai1998,Zhang2001, Troja2007} or late-time accretion onto a nascent BH via fallback of marginally bound ejecta \citep[e.g.][]{Rosswog2007,Cannizzo2011}. 
Unlike the narrow ultra-relativistic prompt jet, this late-time outflow might emerge as a wind with a lower bulk Lorentz factor and a substantially larger opening angle \citep{Bucciantini2012}.

This observational evidence naturally connects compact-object mergers to the growing class of fast X-ray transients (FXRTs). FXRTs are soft X-ray outbursts with durations from minutes to hours, and they often lack an an obvious prompt gamma-ray counterpart \citep{Bauer2017,Xue2019,Yang2019}. 
Their phenomenology (e.g. duration, luminosity, and spectral shape) resembles the features of short GRB X-ray counterparts \citep{Chen2025}. 
It is thus to be expected that some short GRBs (or, more in general, merger-driven GRBs) might be accompanied by an FXRT. 
However, if the relativistic jet is weak or viewed off-axis, gamma-rays can be substantially suppressed while softer, longer-lived X-ray components might remain detectable. 
In such scenarios, the X-ray transient, rather than the gamma-ray spike, could be the primary observable signature of a compact binary merger \citep{Zhang2013}.

\begin{figure}
	\includegraphics[clip, width=0.95\linewidth]{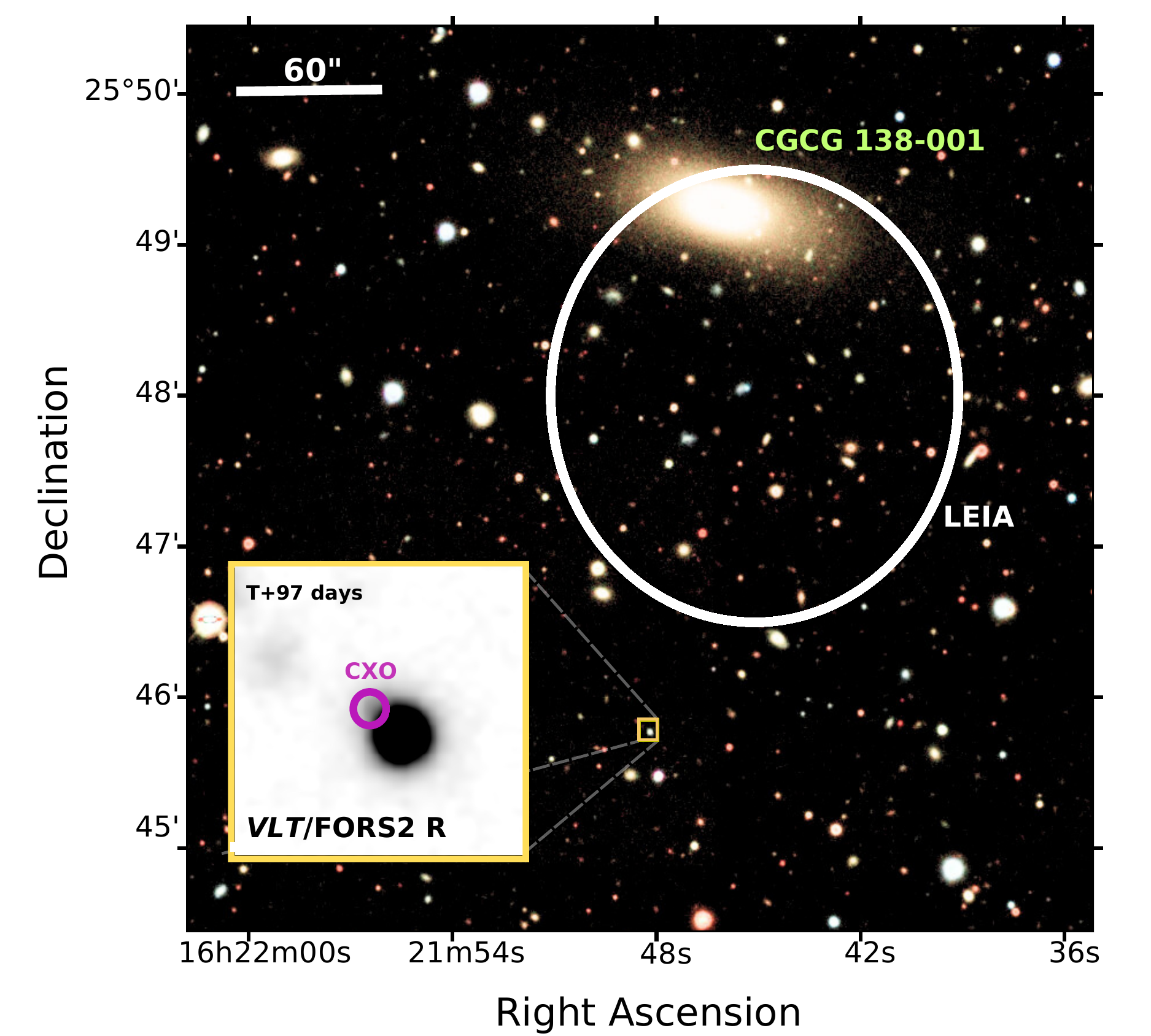}
    \caption{False-colour image of the field of LXT~240402A/GRB~240402B. The preliminary LEIA localisation (white circle) intercepts a bright galaxy, CGCG~138-001, at $\approx$210~Mpc. 
    However, a fading X-ray and optical counterpart was localised further away from it. The inset zooms in on its position. A distant galaxy, partially blended with the nearby star, is seen at the position of the X-ray counterpart.    
    \label{fig:field}}
\end{figure}

The proposed link between compact-object mergers and FXRTs can now be tested, for the first time, thanks to the advent of sensitive wide-field X-ray facilities, such as the \textit{Einstein Probe} (EP), designed to systematically uncover FXRTs and their progenitors \citep[e.g.][]{Yuan2015,Yuan2022}. 
Establishing whether a fraction of FXRTs arise from compact-object mergers would have broad implications for multi-messenger astrophysics, as it would hand us a new tool to probe the diversity of their central engines and outflow structures, while expanding the landscape of EM counterparts of potential GW sources, beyond the standard gamma-ray selection. 

In this work, we present our study of two recent FXRTs - LXT240402A and EP250207b - and explore whether they are the result of a compact-object merger. Without the canonical indicator of a short duration ($\lesssim$2 s) gamma-ray prompt phase, the most immediate discriminator of a merger origin are the properties of the environment: the host-galaxy type, its star-formation rate, and the transient’s location within its host \citep[e.g.][]{Bloom2002,Zhang2007,Fong2013,OConnor2022}. 
Whereas a large fraction of well-localised FXRTs are consistent with being powered by the explosions of massive stars \citep{Sun2025SN,OConnor2025}, both of our targets are potentially associated with two nearby quiescent galaxies with no signs of ongoing star-formation, and are thus prime candidates for merger-driven explosions. 
 
The paper is organised as follows: We present observations of LXT~240402A and EP2502027a in Sect.~\ref{sec:obslxt} and Sect.~\ref{sec:obsEP25027}, respectively. After characterising their prompt emission, afterglow phase, and environment, we place
constraints on their associated kilonova in Sect.~\ref{sec:kn} and discuss other progenitor signatures in Sect.~\ref{sec:other}. 
Throughout this work, we adopt a flat $\Lambda$ Cold Dark Matter cosmology + BAO with $H_0 = 67.66$~$\mathrm{km\ s^{-1}\ Mpc^{-1}}$ and $\Omega_m = 0.3111$ \citep{Planck2020}.
Spectral fits were carried out within {\sc xspec} \citep{Arnaud1996} by minimising the Cash statistics \citep{Cash1979}.  Uncertainties are quoted at the 68\% confidence level unless stated otherwise.

\section{Observations and data analysis} 
\label{sec:obs}

\subsection{LXT~240402A / GRB~240402B}\label{sec:obslxt}
\subsubsection{Prompt Emission}
\label{sec:he}

On April 2, 2024, at 08:47:41 UTC, the FXT LXT~240402A was discovered by the Lobster Eye Imager for Astronomy (LEIA; \citealt{36016}) and localised at RA, Dec (J2000) = 245.438, +25.800 degrees with a 90\% error radius of 1.5\arcmin. 
This localisation intercepts a bright ($r\approx$14 AB mag) nearby ($\approx$210 Mpc) galaxy, CGCG~138-001 (Figure~\ref{fig:field}). 
As discussed in \citet{Dichiara2020}, the probability of a chance alignment between a position with arcmin accuracy and a nearby galaxy is relatively small ($\approx$3\%), which prompts us to consider a possible physical association between LXT~240402A and CGCG 138-001. 
The transient's estimated duration is $\approx$200~s in the 0.5-4~keV band. 

\begin{figure}
	\includegraphics[clip, width=0.95\linewidth]{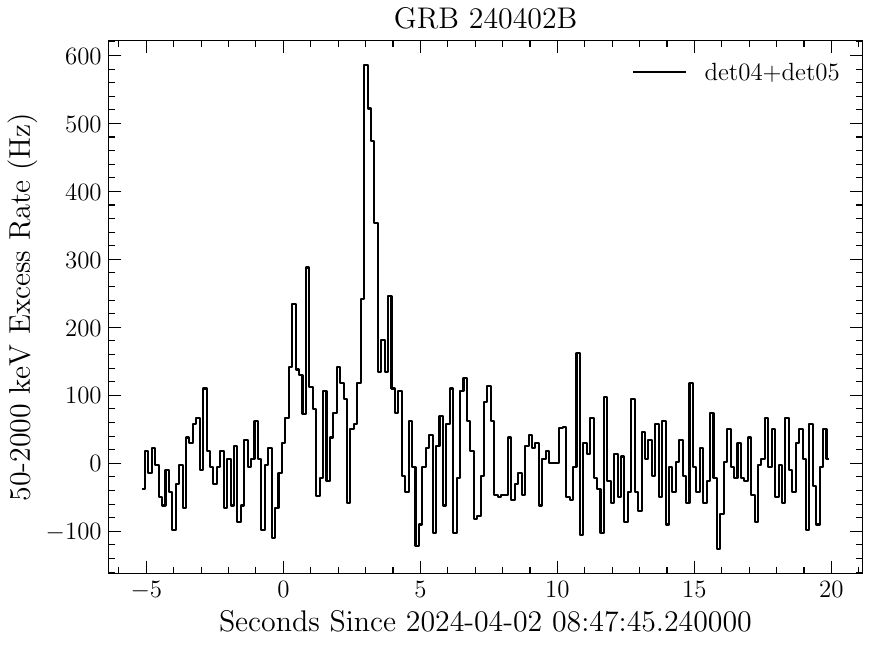}\\
    \vspace{1cm}\\
	\includegraphics[clip, width=0.95\linewidth]{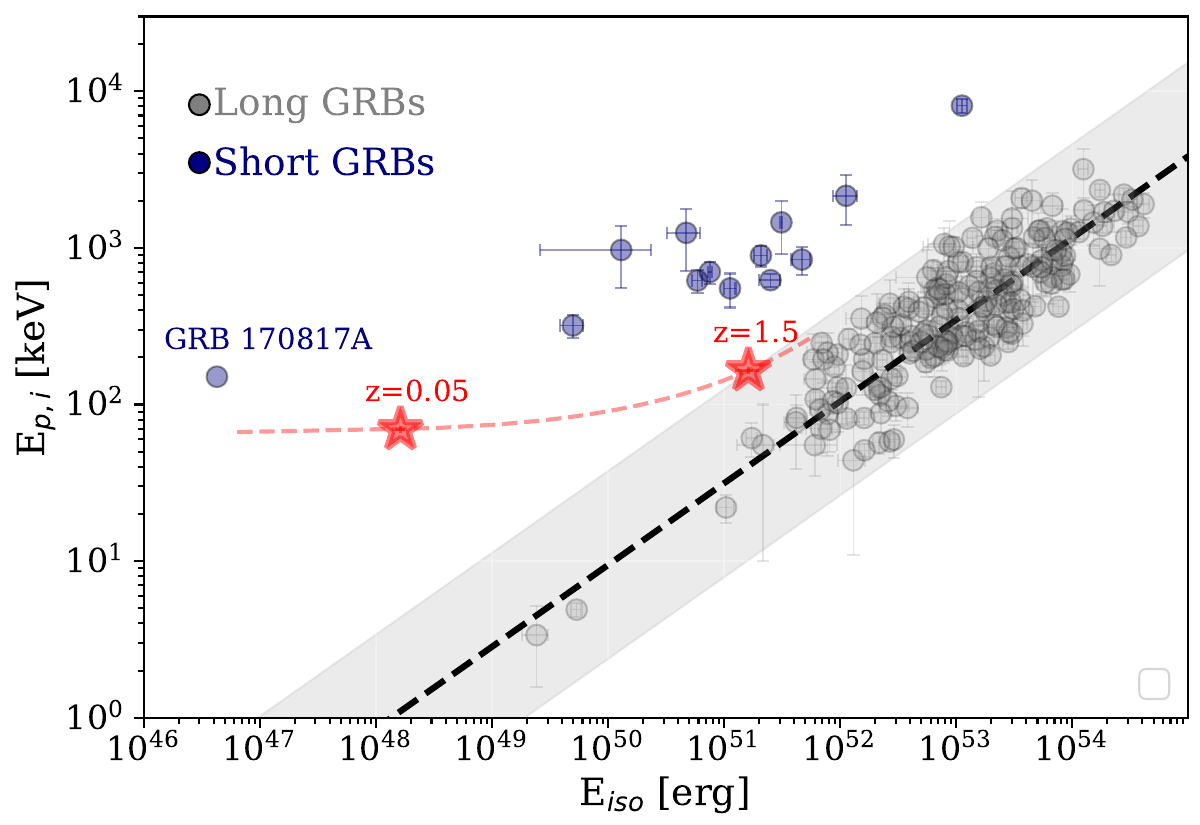}
    \caption{Light curve and $E_{\rm peak}$ - $E_{\rm iso}$ function. Top panel: Background-subtracted 50~keV - 2~MeV light curve of GRB~240402B in 0.125-s bins as observed by Glowbug. The two primary peaks at T0+0.5~s and ~T0+3~s as described in \citet{36030} are visible, with additional sub-structure in the light curve that is overall consistent with that observed by Konus-Wind \citep{36041}.
    Bottom panel: $E_{\rm peak}$ - $E_{\rm iso}$ diagram for long (grey) and short GRBs (blue), updated from \citet{Dichiara2021}. The position of LXT~240402A/GRB~240402B is shown as a function of redshift, highlighting $z=$0.048 and 1.5. The Amati relation (dashed line; \citealt{Amati2008}) and its 2$\sigma$ scatter (grey area) are shown. 
    \label{fig:amati}}
\end{figure}

Simultaneously, GRB~240402B was detected by the Gravitational wave high-energy Electromagnetic Counterpart All-sky Monitor \citep[GECAM-C;][]{Zhang2023,36017}, Konus-Wind \citep{Aptekar1995,36028}, and the Glowbug gamma-ray telescope \citep{Woolf2024,36030}. The gamma-ray emission (Figure~\ref{fig:amati}; top panel) has a relatively short duration, estimated as $T_{90} = 6.6\pm 0.5$~s (15-350~keV), $T_{90}\sim 5.1$~s, and $T_{90}\sim 4.1$~s  for GECAM-C \citep{36017}, Konus-Wind \citep{36028}, and Glowbug \citep{36030}, respectively.
Despite the different durations of the two transients, their temporal and spatial coincidence suggest that GRB~240402B and LXT~240402A are likely connected and produced by the same astrophysical event \citep{36017}. 
Moreover, although the T$_{90}$ values of GRB~240402B slightly exceed the canonical threshold of $\approx$2 s \citep{Kouveliotou1993}, there is a non-negligible probability that it belongs to the population of short GRBs, traditionally driven by NS mergers. 
For example, its duration sits close to the equal-probability threshold ($\approx$5 s) between the long and short classes of the BATSE sample \citep{Donaghy2006}. 
However, since this threshold is instrument-dependent \citep{Bromberg2013}, 
in our case we used the duration distribution of Konus-Wind GRBs \citep{Svinkin2021} to estimate that a burst with $T_{90}\sim 5$~s has a 14\% chance of belonging to the phenomenological population of short bursts.
Taken together, the relatively short gamma-ray duration, its temporally extended X-ray tail, and apparent proximity to a quiescent nearby galaxy make LXT~240402A and GRB~240402B a promising candidate for a compact-object merger origin.

The spectral analysis of the GECAM-C data was reported by \citet{36017} and points to a relatively soft non-thermal spectrum described by a Comptonized model with index of ${-0.62_{-0.14}^{+0.13}}$ and peak energy of $E_\mathrm{peak}={66_{-3}^{+3}}$~keV. Based on these spectral parameters, the total measured fluence is ${7.69_{-0.24}^{+0.25}} \times 10^{-7}$~erg\,{cm}$^{-2}$. 
In Figure~\ref{fig:amati} (bottom panel) we show the position of GRB~240402B in the Amati diagram \citep{Amati2008} as a function of the redshift. 
At the nearby distance of $z \sim 0.048$ ($D_L \approx 210$ Mpc), the inferred isotropic-equivalent gamma-ray energy is  
$E_{\rm iso} \approx 4 \times 10^{48}~\mathrm{erg}$, 
making the burst a clear outlier due to its unusually low energy output. Its position in the $E_p - E_{\rm iso}$ plane fits within the track defined by short GRBs \citep{Zhang2009,Dichiara2021}. 
For comparison, the prototypical merger-driven burst GRB~170817A released $E_{\rm iso} \sim 5 \times 10^{46}$~erg \citep{Abbott2017} and is a clear outlier of both correlations.  
For higher redshifts, $z \gtrsim 1.5$, the isotropic-equivalent energy reaches $E_{\rm iso} \gtrsim 3 \times 10^{52}~\mathrm{erg}$, and the burst becomes  consistent with the standard Amati relation for long GRBs within its $2\sigma$ dispersion (Figure~\ref{fig:amati}).

\begin{table*}
 	\centering
 	\caption{X-rays and optical photometry of LXT~240402A.  \label{tab:observationsLXT240402A}}
\begin{tabular}{cccrccc}
    \hline
    \hline
   \multicolumn{7}{c}{\textbf{X-rays}}  \\
   \hline
  Mid-time & Exposure & Telescope & Instrument &  & Unabsorbed flux&References\\
    (days) & (ks) & &   && (10$^{-14}$ erg cm$^{-2}$ s$^{-1}$)&\\
 \hline
1.53	&	12	&	\textit{EP}	&	FXT	&		&	86.00$\pm$9.00	&	 \citet{gcnep} \\
5.49	&	47	&	\textit{Swift}	&	XRT	&		&	12.97$\pm$8.76	&	 This work, \citet{EvansXRT2009} \\
11.75	&	35	&	\textit{Swift}	&	XRT	&		&	8.64$\pm$4.59	&	This work,  \citet{EvansXRT2009} \\
12.79	&	16	&	\textit{Chandra}	&	ACIS-S 	&		&	9.13$\pm$1.20	&	This work\\
 \hline
 \hline
    \multicolumn{7}{c}{\textbf{Optical/nIR photometry}}  \\
   \hline
   Mid-time & Exposure & Telescope & Instrument & Filter& Magnitude&References\\
    (days) & (s) & &  &   &(AB)&\\
 \hline									
	1.31	&	8100	&	Kinder	&	LOT\&SLT	&	  r 	&	 22.21$\pm$0.10 	&	\citet{36027}\\
	1.93	&	1700	&	VLT	&	FORS2	&	 R 	&	 $22.71\pm0.03$ 	&	\citet{36025}, This work\\
	2.96	&	900	&	VLT	&	FORS2	&	 R 	&	 $23.31\pm0.03$ 	&	This work\\
	4.94	&	900	&	VLT	&	FORS2	&	 R 	&	 $23.76\pm0.04$ 	&	This work\\
	97.73	&	2400	&	VLT	&	FORS2	&	 R 	&	Template 	&	This work\\
\bottomrule
\end{tabular}
\tablefoot{Magnitudes are corrected by Galactic Extinction. Upper limits correspond to a 3$\sigma$ confidence level.}
\end{table*}

\subsubsection{Afterglow phase}

Subsequent follow-up with \textit{Einstein Probe} identified a fading X-ray source at RA, Dec (J2000)= 16:21:48.24, 25:45:46.80 with an uncertainty of 10\arcsec \citep{gcnep}. This position is 2.4\arcmin\ away from the initial LEIA localisation and 3.5\arcmin\ away from CGCG 138-001. 
The large offset between the newly localised X-ray source and its putative host casts doubts about their possible association. Moreover, several extended objects lie close to or within the X-ray position, preventing a secure association with any other galaxy. 

Within the FXT localisation, a potential optical counterpart was identified by \citet{36027}, yielding a precise sub-arcsecond position of RA, Dec (J2000)= 16:21:48.23, 25:45:47.57.
Follow-up observations, performed with the Very Large Telescope (VLT) continued to monitor the source for several days.  
These data were reduced using standard CCD techniques (e.g. bias subtraction, flat-field correction) as implemented in the ESO Reflex environment \citep{Freudling2013}.

\begin{figure}
	\includegraphics[clip, width=0.95\linewidth]{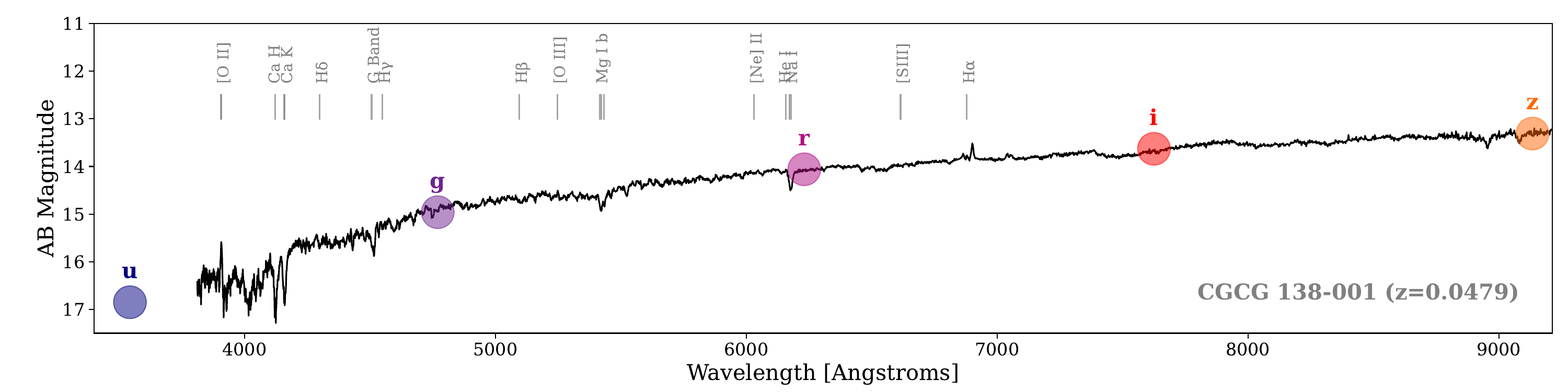}
    \caption{Spectral energy distribution of CGCG 138-001  \citep{sdss2007}. The optical spectrum was renormalised to match the photometric measurements. The most prominent features are identified. 
    \label{fig:CGCG138-001}}
\end{figure}

The photometry of the optical counterpart was complicated by the nearby ($\lesssim$1\arcsec) bright star ($r\sim$20.6 AB) and
by the contribution of an underlying galaxy (Figure~\ref{fig:CGCG138-001}). 
Our refined analysis isolates the transient emission by performing image subtraction with the Saccadic Fast Fourier Transform (\texttt{SFFT}; \citealt{sfft}) software, 
using as a reference a late-time template in $R$ band. 
We then performed PSF photometry on the residual images, calibrating it with the Pan-STARRS PS1 DR2 Catalog \citep{Magnier2020}. 
The detailed photometry for LXT~240402A is listed in Table~\ref{tab:observationsLXT240402A}.

Our analysis shows that, at least for the first few days, the X-ray and optical counterparts fade as a simple power law with temporal index $\alpha_X$=1.00$\pm$0.10 and $\alpha_o$=1.10$\pm$0.01, respectively.

\begin{figure*}
\includegraphics[clip, width=0.95\linewidth]{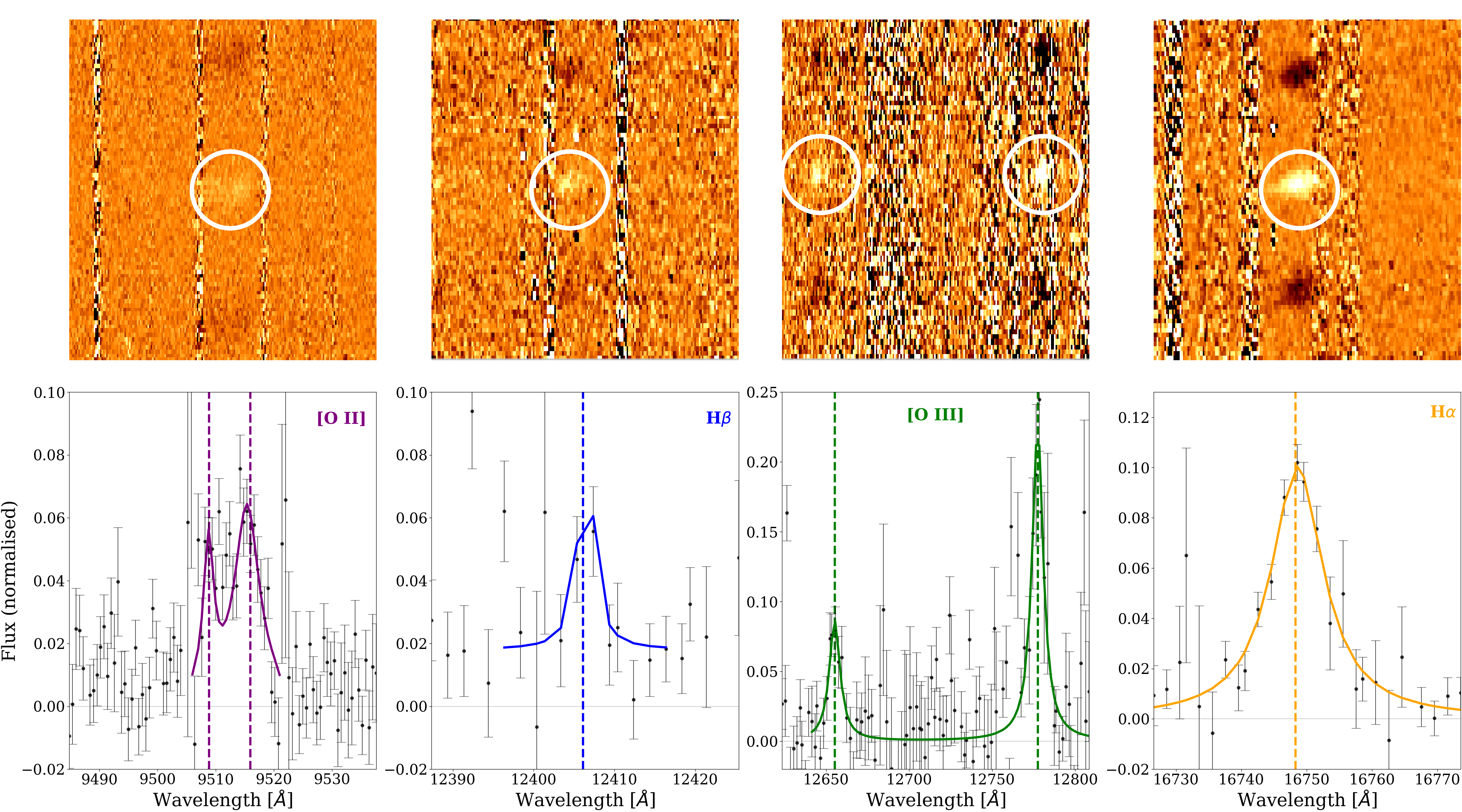}
    \caption{VLT/X-Shooter spectrum of the faint galaxy underlying the position of LXT240402A. The bottom panels show the one-dimensional spectrum (black points), smoothed with a three-pixel binning, along with its standard deviation (black error bars), zoomed in around the identified emission lines. The flux is normalised to its maximum value. The dashed lines mark the positions of the features, while the solid lines show their Lorentzian fits; from left to right: [O II] (purple), $H\beta$ (blue), [O III] (green), and $H\alpha$ (yellow). The top panels display the corresponding emission features in the two-dimensional spectrum (x-axis: wavelength; y-axis: spatial direction), highlighted by white circles.
    \label{fig:specgrb240402b}}
\end{figure*}

\subsubsection{Chandra localisation}

Given the field complexity, an association between the optical and X-ray emission was not immediately evident. Multiple galaxies fall within the FXRT localisation, and optical variability could plausibly arise from nuclear activity (e.g. an AGN) rather than the true FXRT counterpart. 

For this reason, we activated a \textit{Chandra} Discretionary Director's Time observation aimed at precisely localising the X-ray counterpart, thus unambiguously establishing its association with the optical transient. 

We imaged the field using the Advanced CCD Imaging Spectrometer-S (ACIS-S) detector onboard the \textit{Chandra X-ray Observatory} (CXO). The observations began at $T + 12.7$~days with a total exposure time of 16.4~ks (ObsID: 29380; PI: Troja).
Data were reprocessed and analysed using the \texttt{CIAO} software package \citep[\mbox{v. 4.16;}][]{Fruscione2006} and the calibration database files (\texttt{caldb}; v. 4.11.5). Absolute astrometry was corrected using four common sources with the PS1 catalogue \citep{Magnier2020}. 

Within the FXRT localisation, a single X-ray source
is detected at RA (J2000) = 16$^{\mathrm{h}}$21$^{\mathrm{m}}$48.25$^{\mathrm{s}}$, Dec (J2000)=+25$^\circ$45$^\prime$47.18$^{\prime\prime}$
with a 1\,$\sigma$ positional uncertainty of 0.3\arcsec. This position is consistent with the optical localisation and confirms its association with the X-ray afterglow. 

The source brightness was estimated by performing aperture photometry using a circular source region with a radius of 1.5\arcsec\ and estimating the background from a source-free concentric annulus with inner and outer radii of 6\arcsec\ and 25\arcsec, respectively. 
We measured a total of 65.9 net counts and, after correcting for point spread function (PSF) losses, we derived a net count rate of ${(4.4 \pm 0.5) \times 10^{-3}\,\mathrm{cts}\,\mathrm{s}^{-1}}$ in the 0.5-8~keV energy band. 
This was converted into flux by fitting the spectrum with an absorbed power-law (model \texttt{tbabs$\times$pegpwrlw}). 
Based on the best fit model with ${N_{\mathrm{H}} = 4.35 \times 10^{20} \mathrm{~cm}^{-2}}$ \citep{Willingale2013} and photon index of $\Gamma_{\mathrm{X}} = 1.7\pm0.4$, we derive an unabsorbed flux $F_{\mathrm{X}} = (9.3^{+1.2}_{-1.1}) \times 10^{-14}\ \mathrm{erg\ s^{-1}\ cm^{-2}}$ in the 0.3-10~keV band.

\subsubsection{Environment}
\label{sec:host}

The radio galaxy CGCG 138-001 at a redshift of $z = 0.04798 \pm 0.00001$ \citep{sdss2007} (see Figure~\ref{fig:CGCG138-001}) was initially proposed as a potential host of LXT~240402A \citep{36016}, based on its spatial proximity to the LEIA localisation and its relatively close distance. 
Subsequent optical and X-ray counterpart localisations shifted the position away from this system, yet on probabilistic grounds CGCG 138-001 remains a plausible host candidate with $P_{cc}\approx$3\% \citep{Dichiara2020}.

The galaxy exhibits a smooth spheroidal and slightly disturbed morphology, with no signs of ongoing star formation or disc-like substructures (Figure~\ref{fig:field}). Its spectrum displays a red continuum with strong absorption features detected in the optical band (Figure~\ref{fig:CGCG138-001}), including the Calcium H and K lines (Ca II ($\lambda$3933, 3968), the G band (CH $\lambda$4300), the Sodium D doublet (Na I $\lambda$5889, 5895) and the Magnesium triplet (Mg I b $\lambda$5167, 5173, 5184), indicating an evolved and metal-rich stellar population (see Figure~\ref{fig:CGCG138-001}). Such an old and massive stellar population provides the ideal conditions for the formation of compact-object binaries capable of producing a merger \citep[e.g.][]{Mapelli2018}.

The offset  between the galaxy and the transient's position is significant. An angular separation of $\approx$3.5\arcmin\ corresponds to a projected physical offset of $\approx$220~kpc between the FXRT progenitor and its host. This is higher than any other offset derived so far from short GRBs
\citep{Bloom2006,Troja2007,Tunnicliffe2014,Fong2013,OConnor2022}, and at the upper end of the offset distribution predicted by progenitor models \citep{Fryer1997, Behroozi2014,Beniamini2016}.
However, when renormalised for the galaxy's half-light radius, $r_h\sim16$\,\arcsec\ \citep{Dey2019}, it corresponds to $\approx$13$r_e$ which is comparable to highly offset short bursts, such as GRB070809 and GRB090515 \citep{Zevin2020}. 
Whereas the latter were explained by invoking a large  ($\gtrsim$200 km\,s$^{-1}$) natal kick of the progenitor, 
in this case we must consider that CGCG 138-001 resides within a small ($N\approx$6) galaxy group \citep{Smith2012, Tempel2017}. In such an environment, gravitational interactions and ram pressure stripping \citep{GunGott1972,Hester2006} may have influenced the galaxy's morphology and displaced the burst progenitor system from its birthplace on scales of tens to hundreds of kiloparsecs \citep{Zemp2009,Dichiara2025}, making a large natal kick unnecessary to explain the observed offset.
Based on the properties of the galaxy and its environment, the link between LXT240402A and CGCG 138-001 remains physically plausible. 

Alternatively, we consider the possibility that the transient originated within a faint underlying galaxy, visible in the late-time imaging (Figure~\ref{fig:field}). 
Spectroscopic observations of this galaxy were undertaken with the VLT/X-Shooter spectrograph \citep{Vernet2011} at $\approx$30 d after the burst, at an average airmass of 1.8 and in good seeing conditions (seeing $\sim 0.6$\arcsec). 
The spectrum consists of $4\times 1200$s exposures and covers the wavelength range 3900–21000 \AA. 
After applying standard prescriptions for data reduction and calibration within the ESO-Reflex pipeline \citep{Modigliani2010}, we identify a clear set of nebular emission features (Figure~\ref{fig:specgrb240402b}), including the [O II] $\lambda\lambda$3726,3729 doublet, [O III] $\lambda\lambda$4959,5007, and the Balmer lines H$\alpha$ and H$\beta$. Using the H$\alpha$ spectral lines and the OIII doublet, a Gaussian fit of the absorption lines was performed using \texttt{specutils}, yielding a redshift $z=1.55130 \pm 0.00011$
consistent with preliminary analysis reported in \citep{36385}.
Assuming negligible extinction based on the ratio of H$\alpha$ and H$\beta$, we estimate the galaxy's star formation rate as SFR$\approx$\,5.4$\times$10$^{-42}$\,$L_{H\alpha}$\,$\gtrsim$\,15 $M_{\odot}$ yr$^{-1}$\citep{Kennicutt2012}. 

From the late-time optical template, we derive a galaxy brightness of $R\sim\!23.9$ AB mag
and a half-light radius $r_h$\,$\sim$0.3\arcsec. 
We follow the standard formalism of \citep{Bloom2002} to estimate the probability of a chance alignment $P_{cc}\,\approx$0.5\%. 
This value is lower - though only by a factor of a few - than the chance-alignment probability with the nearby galaxy. Whereas this favours the distant galaxy as the most likely host galaxy, our prior knowledge of GRBs and their broad offset distribution does not allow a definitive conclusion simply based on positional arguments. 

\subsection{EP250207b} 
\label{sec:obsEP25027}

\subsubsection{Prompt emission}
\label{sec:heEP25027b}

\begin{figure}
	\includegraphics[clip, width=0.95\linewidth]{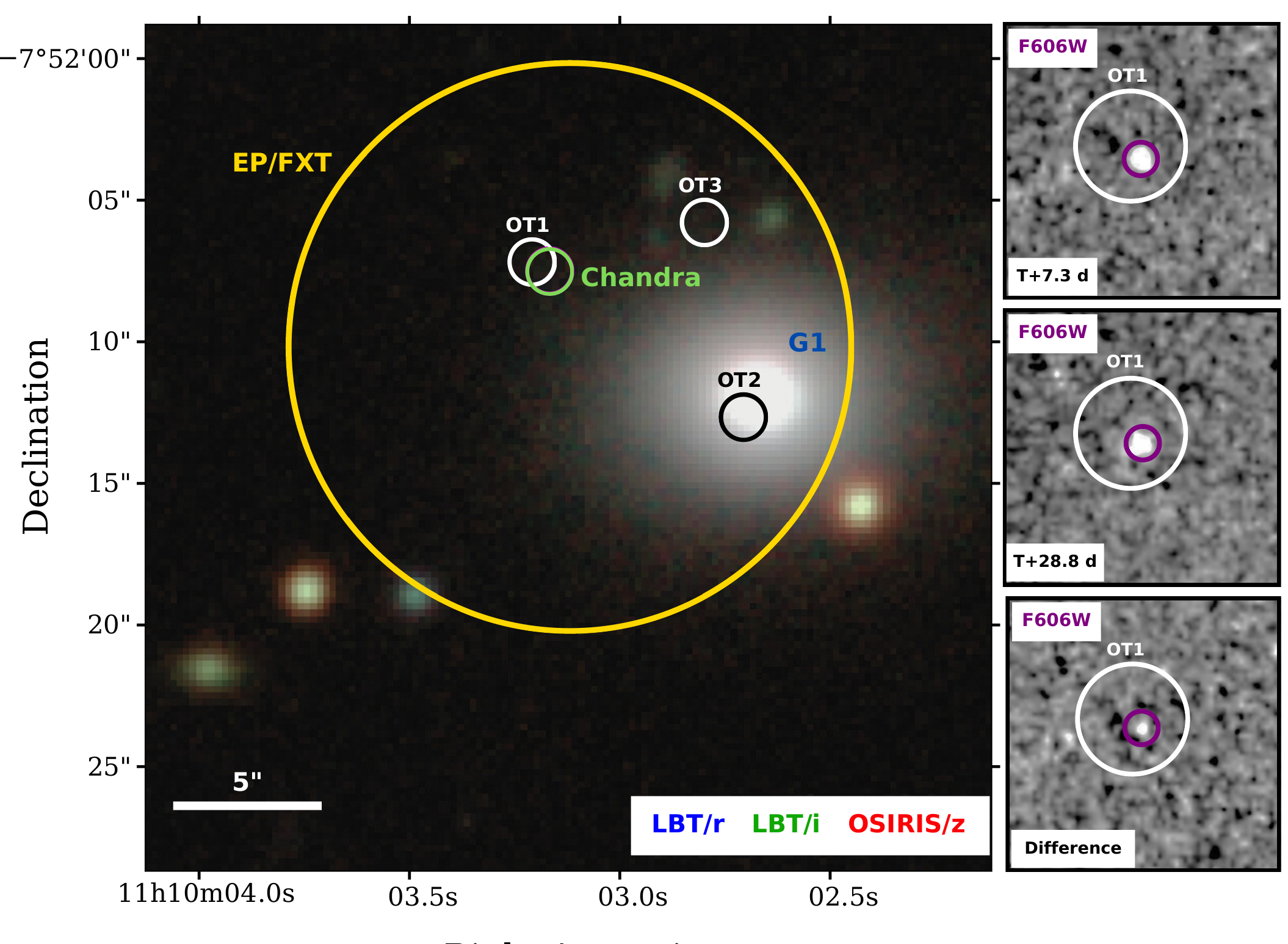}
    \caption{False-colour image of the field of EP250207b. Within the EP/FXT error circle (yellow), three optical candidate counterparts were reported by NOT (OT1; \citealt{39300}), Gemini (OT2;\citealt{39287}), and the Liverpool Telescope (OT3; \citealt{39281}), respectively. The precise position by \textit{Chandra} links EP250207b with OT1, located $\approx$9.3\arcsec away from the centre of the galaxy G1. 
    The insets show the two epochs of $HST$ optical observations, and their difference, which confirms fading. 
    The purple circle shows the effective radius, $R\approx$0.15\arcsec, used for deriving the probability of chance alignment. 
    \label{fig:fieldEP250207b}}
\end{figure}

EP250207b was detected by the Wide-field X-ray Telescope (WXT) on board the Einstein Probe (EP) mission on $T_0 = $ 2025 February 7 21:47:56 UTC and localised by the EP/FXT at RA, Dec (J2000)= 11:10:03.12, -07:52:10.20 with an uncertainty of 10\arcsec \citep{39266}. The error circle encompasses a bright galaxy (hereafter G1) at $z=0.082$ (Figure~\ref{fig:fieldEP250207b}), raising the possibility of a physical association between the FXRT and the nearby galaxy. 

\begin{figure}
	\includegraphics[clip, width=0.48\textwidth]{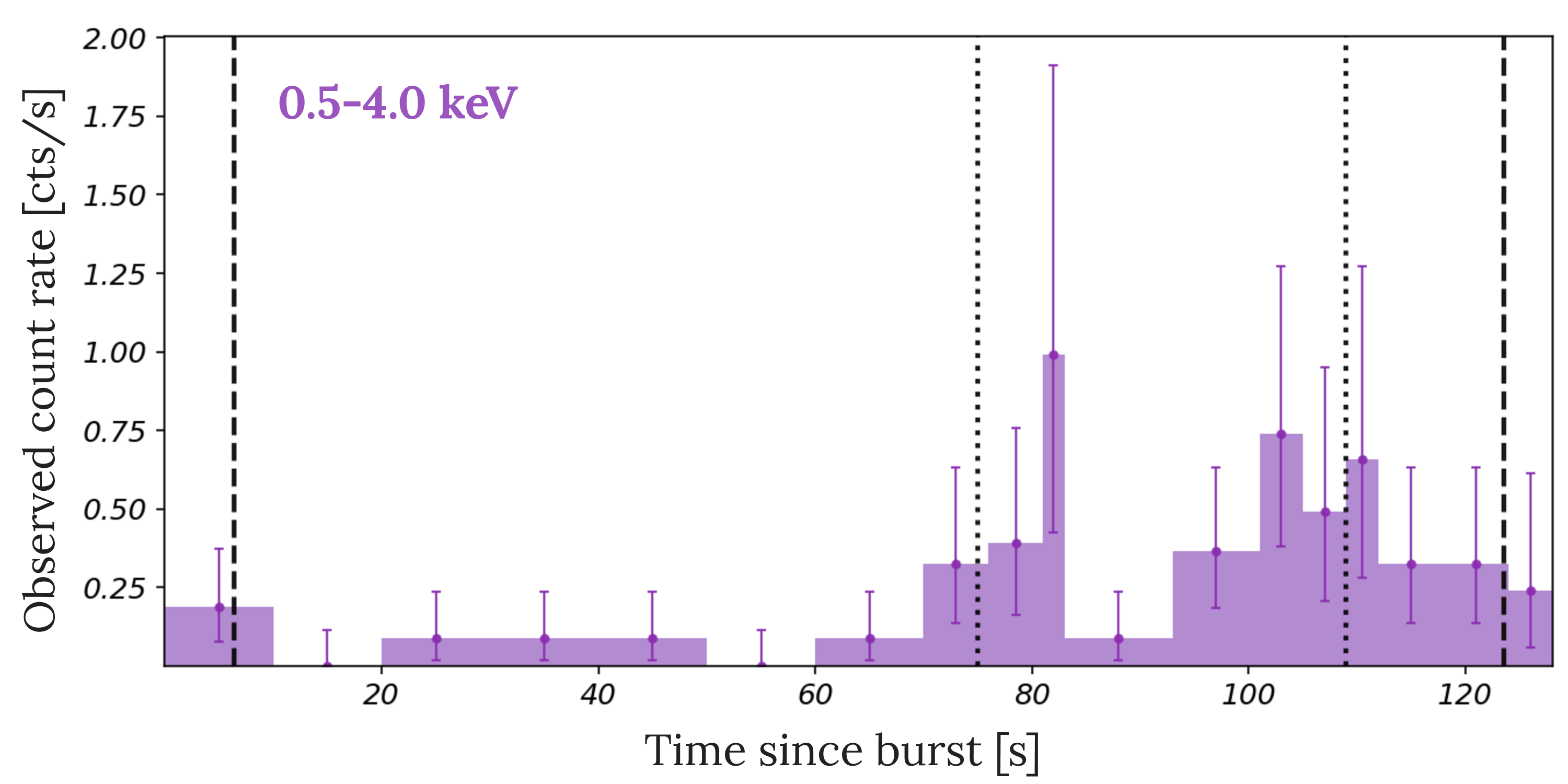}
    \caption{EP/WXT light curve of EP250207b in the 0.5-4.0~keV band,
    adaptively binned to achieve either a 3$\sigma$ significance per bin or a maximum bin size of 10 s. Error bars are at the 68\% confidence level. The vertical dashed (dotted) lines mark the 
    $T_{90}$ ($T_{50}$) interval. 
    \label{fig:lcEP}}
\end{figure}

All EP data were reduced and calibrated following standard techniques implemented in the WXT Data Analysis Software (\texttt{WXTDAS}) and FXT Data Analysis Software (\texttt{FXTDAS}, v.1.2).
The WXT lightcurve (Figure~\ref{fig:lcEP}) was derived by extracting the source counts from a circular region with a standard 9\arcmin\ radius. The background contribution (on average $\approx$0.01 cts s$^{-1}$) was estimated from a concentric source-free annular region with radii of 18 arcmin and 36 arcmin. The source is characterised by low level emission from $T_0$ to $T_0 + 50$ s (5 total counts, $\approx 3.5 \sigma$ significance; \citealt{KraftBurrowsNousek90}), followed by a lull then a main multipeaked outburst from $T_0 + 70$~s to $T_0 + 125$~s (22 total counts), when the observation of the field ended. 
The observed variability ($\Delta\,t\,/\,t\,\lesssim$0.1) is consistent with prompt emission and X-ray flares rather than afterglow onset, and points to a long-lived central engine. 
Using the standard definition \citep{Kouveliotou1993} of $T_{90}$ ($T_{50}$) as the time interval over which 90\% (50\%) of the total net counts is measured, we derive $T_{90}$=117 s and $T_{50}$=34 s in the 0.5-4~keV band. 
Because the WXT observation was interrupted during the outburst, these values are lower bounds on the true durations. 

The spectrum over the time interval of the main burst (from $T_0 + 70$~s to $T_0 + 125$~s) favours an absorbed power-law model (model \texttt{tbabs$\times$pegpwrlw}, CSTAT=14 for 21 degrees of freedom, dof) with $N_H$=4.2$\times10^{20}$ cm$^{-2}$ \citep{Willingale2013} and photon index $\Gamma$=0.4$\pm$0.5 over a blackbody (CSTAT=15.6 for 21 dof). 
The best fit model yields an unabsorbed X-ray flux 
of (1.3$\pm$0.5)$\times$10$^{-9}$ erg cm$^{-2}$ s$^{-1}$ and a total fluence of (9.0$\pm$0.3)$\times$10$^{-8}$ erg cm$^{-2}$. 

\begin{figure}
    \centering
    \includegraphics[width=1\linewidth]{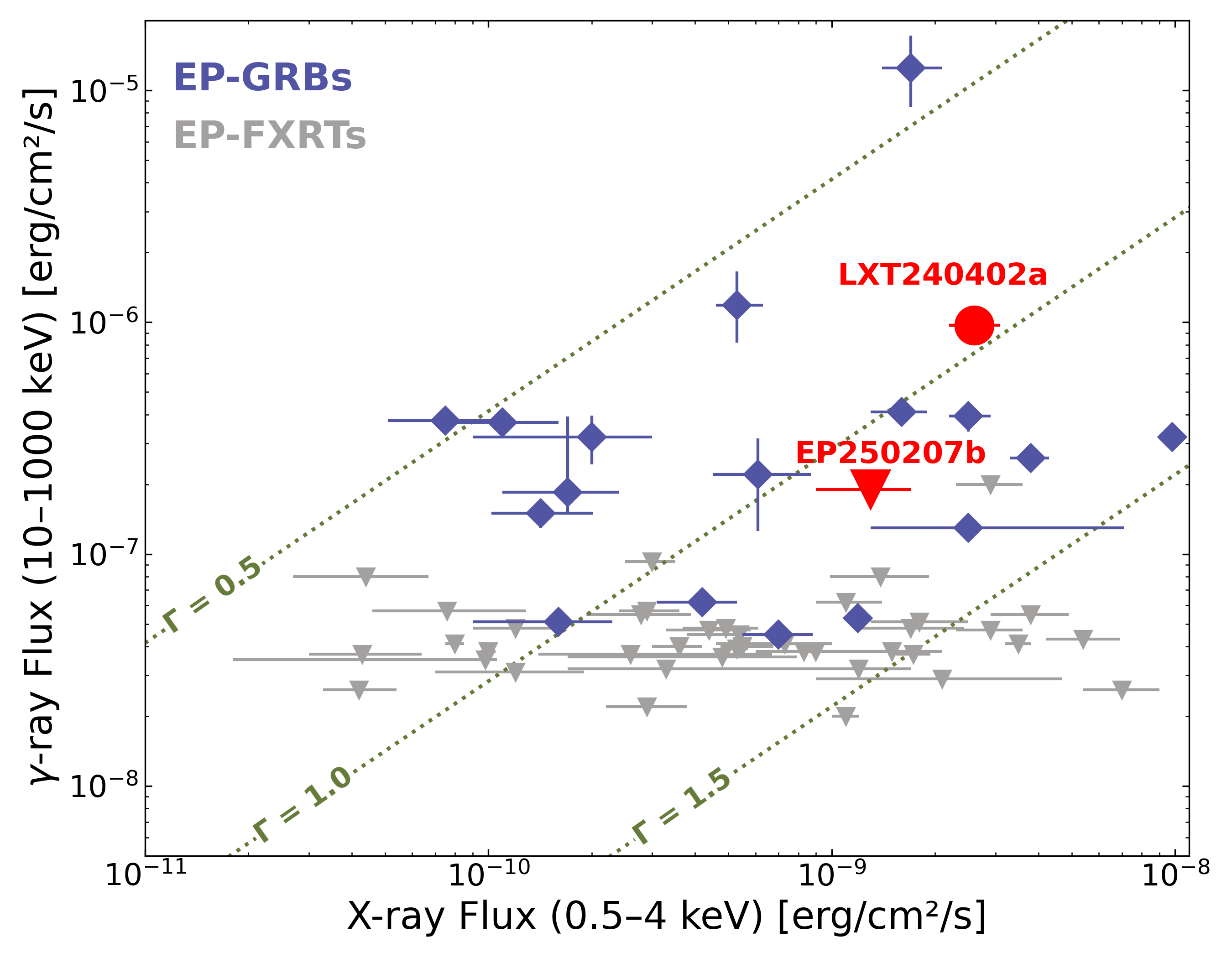}
    \caption{Observed gamma-ray flux (10-1000~keV) versus X-ray flux (0.5-4~keV) for EP-discovered FXRTs, updated from \citet{Yadav2025}. 
    LXT 240402a and EP250207b are highlighted. 
    Dashed lines represent power-law spectra with photon indices $\Gamma = 0.5$, $1.0$, and $1.5$.}
    \label{fig:xgamma}
\end{figure}

No gamma-ray counterpart was associated with EP250207b. The burst was Earth-occulted for \textit{Fermi} Gamma-ray Burst Monitor and outside the field of view of the \textit{Swift} Burst Alert Telescope. 
At the time of the explosion, Konus-Wind was observing the whole sky. Using waiting-mode data within the interval $T_0\!\pm\!1000$~s,
we found no significant ($>$5$\sigma$) excess over the background in S1 KW detector, with the smallest incident angle, on temporal scales from 2.944~s to 1000~s.
We estimate an upper limit (90\% c.l.) on the 10-1000~keV peak flux to $1.9\times10^{-7}$ erg\,cm$^{-2}$\,s$^{-1}$ for a typical GRB spectrum, described as a Band function \citep{Band1993} with low-energy index $\alpha=\,-1$, high-energy index $\beta=\,-2.5$, and spectral peak $E_\mathrm{peak}=300$~keV) on a 2.944~s timescale.

Based on these constraints, EP250207b is consistent with both the population of GRBs and gamma-ray dark FXRTs detected by EP (Figure~\ref{fig:xgamma}).

\begin{figure}
	\includegraphics[clip, width=0.95\linewidth]{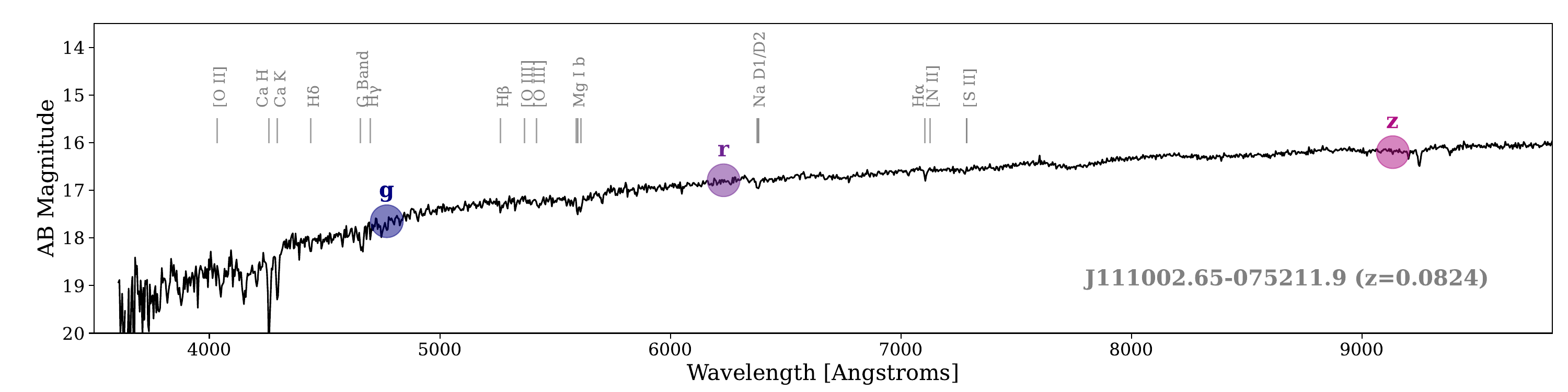}
    \caption{Spectral energy distribution of the galaxy G1 (see Figure~\ref{fig:fieldEP250207b}) from the DESI catalogue \citep{desi}
    \label{fig:spectrumgalaxy250207b}}
\end{figure}

\begin{table*}
 	\centering
 	\caption{X-rays, optical, and radio photometry of EP250207b.}
    \label{tab:observationsEP250207b}
\begin{tabular}{lcccrc}
    \hline
    \hline
   \multicolumn{6}{c}{\textbf{X-rays}}  \\
   \hline
Mid-time& Exposure & Telescope & Instrument & Band & Unabsorbed flux\\
(days) & (ks)& & & (keV) & (10$^{-14}$ erg cm$^{-2}$ s$^{-1}$) \\

 \hline
0.71	&	3	&	\textit{EP}	&	FXT	&	0.3-10	&	28.0$\pm$5.0	\\	
1.86	&	5	&	\textit{EP}	&	FXT	&	0.3-10	&	4.0$\pm$1.0	\\	
2.5     &   9   &   \textit{EP} &  FXT  &   0.3-10  &   3.2$\pm$0.7	\\
4.90	&	24	&	\textit{Chandra}	& ACIS-S &	0.3-10  &	  0.4$\pm$0.2	\\	
 \hline
 \hline
    \multicolumn{6}{c}{\textbf{Optical/nIR}}  \\
   \hline
Mid-time& Exposure & Telescope & Instrument & Filter& Magnitude \\
(days) & (s)& & &  & (AB) \\
 \hline			

4.31	&	480	&	GTC	&	OSIRIS	&	$z$	&	$>23.8$	\\
7.42    &  2020 &  HST\footnote{The magnitudes quoted in the table were obtained via image subtraction \citep{sfft}. The values derived from aperture photometry are (AB mag):  
$F606W\!=\!26.41\!\pm\!0.09$, $F105W\!=\!25.75\!\pm\!0.13$, $F125W\!=\!26.83\!\pm\!0.08$, $F160W\!=\!25.82\!\pm\!0.11$. In the second epoch, we estimate $F606W\!=\!26.90\!\pm\!0.10$, $F105W\!=\!26.59\!\pm\!0.16$, $F125W\!=\!26.67\!\pm\!0.19$, $F160W\!=\!26.29\!\pm\!0.16$.
} & WFC3  &      $F606W$ & 28.0$\pm$0.5 \\
7.42    &  2212 &  HST & WFC3  &      $F105W$ & 26.6$\pm$0.2 \\
7.42    &  2212 &  HST & WFC3  &      $F125W$ & 26.8$\pm$0.3 \\
8.74    &  2212 &  HST & WFC3  &      $F160W$ & 27.1$\pm$0.3 \\
8.66	&	819	&	GTC	&	EMIR	&	$J$	&	$>23.5$	\\ 
8.70	&	1461	&	GTC	&	EMIR	&	$K_s$	&	$>23.7$\\
9.16	&	462	&	GTC	&	EMIR	&	$H$	&	$>24.0$	\\ 
11.27	&	1404	&	GTC	&	EMIR	&	$K_s$	&	$>23.2$	\\
12.48	&	1800	&	LBT	&	LBC	&	 $r$ 	&	$>25.8$	\\
12.48	&	1800	&	LBT	&	LBC	&	 $i$ 	&	$>24.9$	\\
23.26	&	1800	&	VLT	&	FORS2	&	i	&	$>25.1$		\\
23.28	&	1800	&	VLT	&	HAWKI	&	$K_s$	&	$>23.8$	\\

 \hline
 \hline
    \multicolumn{6}{c}{\textbf{Radio}}  \\
   \hline
Mid-time& Exposure & Telescope & Array & Band& Flux Density\\
(days) & (hr)& & &  & ($\mu$Jy) \\
 \hline			

4.40	&	0.50	&	VLA 	&		A&	6.0~GHz	&	$<\!21$	\\
7.41	&	0.67	&	VLA 	&	A -> D	&	6.0~GHz	&	$<\!174$	\\
11.78	&	6.65	&	ATCA 	&	6D	&	5.5~GHz	&	$<\!39$	 \\
11.78	&	6.65	&	ATCA 	&		6D	&	9.0~GHz	&	$<\!27$	\\
38.54	&	3.92	&	ATCA 	&		6D	&	5.5~GHz	&	$<\!45$	\\
38.54	&	3.92	&	ATCA 	&		6D	&	9.0~GHz	&	$<\!42$	\\

\bottomrule
\end{tabular}
\tablefoot{Magnitudes are corrected by Galactic Extinction. Upper limits correspond to a 3$\sigma$ confidence level.}
\end{table*}

\subsubsection{Afterglow phase}

EP/FXT began follow-up observations of EP250207b at T+17~h for a total of 3~ks and localised its X-ray afterglow at RA, Dec (J2000)= 11:10:03.12, -07:52:10.20 with an uncertainty of 10\arcsec~\citep{39266}. 
Subsequent visits at T+1.8~d and T+2.5~d, with an exposure of 5~ks and 9~ks respectively, confirmed rapid fading of the counterpart.

The time-averaged X-ray spectrum (from T+17 h to 2.5 d) can be fitted with an absorbed power law with a fixed Galactic equivalent hydrogen column density of $4.2\times 10^{20}$~cm$^{-2}$ and a photon index of $\Gamma=1.9_{-0.2}^{+0.2}$ (C-STAT= 139 for 138 dof).
This is much softer than the value inferred from the WXT observations and is consistent with typical afterglow spectra \citep{EvansXRT2009}.  

The afterglow properties were constrained in the radio band using the Very Large Array (VLA) in C-band with centre frequency of 6 GHz and bandwidth of 4~GHz, and the Australian Telescope Compact Array (ATCA) in C- and X-band at the central frequencies of 5.5 and 9 GHz with the bandwidth of 2~GHz per band.
The VLA data were flagged, calibrated and imaged in CASA \citep{CASA2022} using standard procedures. The VLA (ATCA) primary and bandpass calibrator was 3C286 (1934-638), and the phase calibrator was J1130-1449 (1128-047). 
The rms noise values in the final (cleaned and restored) images were evaluated in regions away from  bright side-sources.
The results are presented in Table~\ref{tab:observationsEP250207b} where upper limits are given at the 3 $\sigma$ level.

Given the potential association with a nearby galaxy, deep optical and near-infrared imaging was carried out to ascertain the presence of a kilonova. 
Unfortunately, our follow-up was delayed by the late announcement of the transient, reported to the public 2.5 days after the burst \citep{39266}. 
We observed the field using the VLT equipped with the FORS2 and HAWK-I cameras, the Large Binocular Telescope (LBT) with its optical camera LBC, and the Gran Canarias Telescope (GTC) equipped with the OSIRIS and EMIR cameras. 
Data were reduced following standard CCD techniques implemented in the official pipelines
\citep{Freudling2013,Pascual2010} and the source brightness was estimated using aperture photometry calibrated against the PanStarrs PS1 DR2 \citep{Flewelling2018} and 2MASS \citep{2MASS} catalogs. 

In addition, two epochs of \textit{Hubble Space Telescope} (Program ID 17806; PI: Tanvir; \citealt{Jonker2025}) were undertaken at $\approx$7.5~d and 
$\approx$29 d to observe this fast X-ray transient.
We retrieved the pre-processed images from the public archive, aligned them to within one pixel using \textsc{Tweakreg} and corrected them using \textsc{AstroDrizzle} to a final pixel scale of 0.06\arcsec/pixel for the IR filters, and 0.02 \arcsec/pixel for the $F606W$ filter. 
The log of observations is reported in Table~\ref{tab:observationsEP250207b}.

\subsubsection{Chandra localisation}

Three potential optical counterparts were initially identified within the FXRT localisation and reported by the 
Nordic Optical Telescope (OT1; \citealt{39300}),  
the Liverpool Telescope (OT2; \citealt{39281}), and
\textit{Gemini North}/GMOS (OT3; \citealt{39287}).

For this reason, we activated a \textit{Chandra} Discretionary Director's Time observation aimed at precisely localising the X-ray afterglow and unambiguously identifying the true optical counterpart. \textit{Chandra} observations (ObsID: 30797; PI: Troja) began on 2025 Feb 12.8 UT ($T+4.9$~days after the trigger), using the ACIS-S camera for a total exposure of 24.7~ks. 

Within the FXRT localisation, we detect ($\gtrsim$5\,$\sigma$ significance) a single X-ray source with a total of 5 counts within a 1.5\arcsec\ aperture. 
We improved the native astrometry by using four common sources in the PS1 DR2\footnote{\url{http://panstarrs.stsci.edu/}} \citep{Flewelling2018} and obtained a refined X-ray position of RA, Dec (J2000) = 11:10:03.186, -07:52:07.52 with a 1\,$\sigma$ uncertainty of 0.5 arcsec. 
Assuming an absorbed power-law spectrum with a photon index of 2 and a hydrogen column density of $4.24 \times 10^{20}$ cm$^{-2}$, we derive an unabsorbed X-ray flux of 4.5$^{+2.6}_{-2.0}\times10^{-15}$\,erg\,cm$^{-2}$\,s$^{-1}$.  

Our \textit{Chandra} observation firmly establishes the connection between the FXRT and the optical counterpart OT1 (Figure~\ref{fig:fieldEP250207b}). 
The combined EP-\textit{Chandra} X-ray lightcurve shows a fast power-law decay with slope $1.91\pm$0.18, consistent with a post jet-break phase \citep{vanEerten2012}.

\begin{figure*}
	\includegraphics[clip, width=0.47\textwidth]{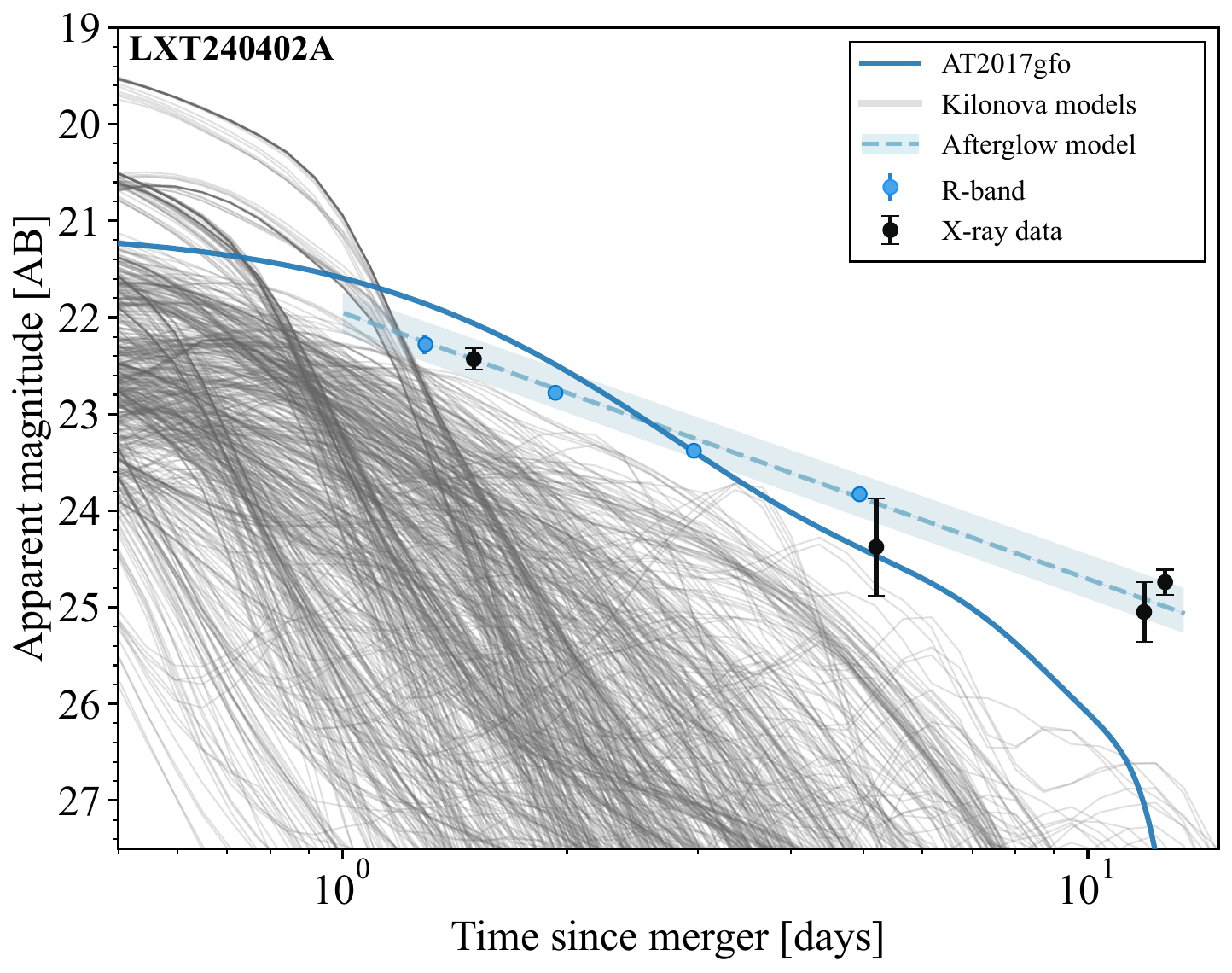}
    \hspace{0.3cm}
    \includegraphics[clip, width=0.47\textwidth]{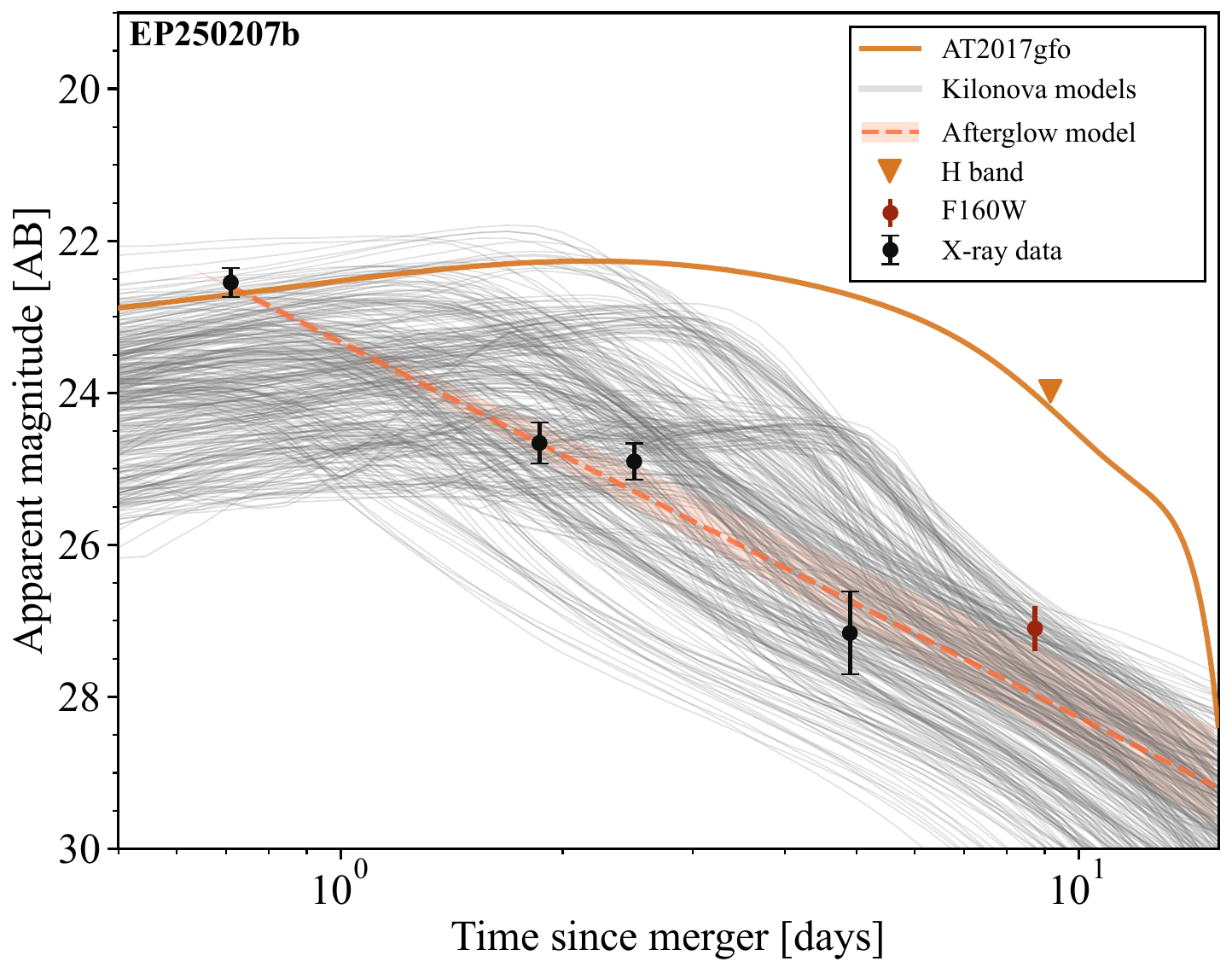}
    \caption{Light curve comparison between models and observations of LXT240402B at $z\sim$0.048 (optical; left) and EP250207b at $z\sim$0.082 (near-infrared; right).  The thick solid line shows the light curve of AT2017gfo at the same redshift. Detections (upper limits) are shown by circles (downward triangles). 
    The dashed line traces the afterglow model, extrapolated from X-ray energies,  with its $1\,\sigma$ uncertainty band. X-ray data were rescaled for plotting purposes. 
    In gray we report the ensemble of kilonova models consistent with the observational constraints. 
    \label{fig:kn}}
\end{figure*}

\subsubsection{Environment}

The location of EP250702b intercepts the outskirts of the bright galaxy G1 at $z\!=\!0.0824$. 
From our GTC/EMIR images, we estimate $H$=$15.89\pm0.03$ AB from the MAG\_AUTO values
in Source Extractor \citep{Bertin1996}, calibrated against 2MASS \citep{2MASS}. 
From the $HST$ $F606W$ image we derive the transient's position RA, Dec (J2000) = 11:10:03.206, -07:52:07.32  with a 1\,$\sigma$ uncertainty of 0.014\arcsec, within the \textit{Chandra} localisation.
We measure a galacto-centric offset of $\approx$9.3\arcsec\ which translates into a projected physical offset of $\approx$14.4~kpc at $z=0.082$.
Based on the observed galaxy number counts in $H$ band \citep{Windhorst2011} and projected angular offset, the probability of a chance alignment between EP250702b and the nearby galaxy G1 (Figure~\ref{fig:fieldEP250207b}) is $P_{cc}\,\approx$0.5\%, making it a likely host galaxy candidate.

The galaxy's spectrum from the DESI catalog \citep{desi}  displays a red continuum with multiple absorption features - most prominently Ca\,\textsc{ii} H\&K ($\lambda\lambda\,3934,3969$) and the G band ($\lambda\,4305$) - and no nebular lines (Figure~\ref{fig:spectrumgalaxy250207b}). This is typical of an early-type galaxy with an old stellar population and no ongoing star formation. 
To characterise the galaxy morphology, we model our VLT $i$-band image with GALFIT \citep{Peng2002} using PSF‐convolved, two–dimensional light profiles. 
The galaxy's surface brightness can be modelled by a Sersic profile with characteristic half-light radius $r_h\,\approx$3.5 kpc and index $n\approx$1.94, indicating a lenticular morphology. Residual maps show no strong non-axisymmetric structures, and adding extra components does not yield a significant improvement of the fit.
Based on this analysis, the host normalised offset is $\approx$4.2$r_h$.

If EP250207b is associated with G1, the surrounding environment favours an old progenitor system, consistent with a compact binary merger. At $z\approx0.082$, the Konus Wind flux upper limit (Section~\ref{sec:heEP25027b}) would correspond to $L_{\gamma, \rm{iso}} \lesssim\!4\times10^{48}$ erg s$^{-1}$ (10-1000 keV), which is orders of magnitude lower than cosmological short GRBs.
However, as with LXT240402A, a more distant host galaxy remains a viable alternative. 
\textit{HST} imaging  shows a flattening of the optical/nIR light curve between 7 and 30~d, a behaviour not expected from standard kilonova models. This plateau can be interpreted either as the contribution of a faint, underlying background galaxy or as the emergence of a supernova component.

In the former case, the probability of a chance alignment between the transient and a faint galaxy is comparable to the probability of association with G1. The first \textit{HST} epoch localises the transient with mas precision. Assuming that the second epoch is dominated by host-galaxy light, we estimate \(r_h \approx 0.07\arcsec\).
We then derive the chance probability in a standard fashion as $P_{cc}=1-{\rm exp}(\pi R^2 \sigma)$ where
$R\sim\,2\,r_h$ and $\sigma$ is the surface density of a galaxy brighter than $H\lesssim$26.3 AB mag \citep{Windhorst2011}. 
The resulting value is $P_{cc}\approx$0.6\%, broadly equivalent to the case of G1.

Because a third late-time epoch is not available yet, we cannot rule out that the second \textit{HST} epoch still contains significant transient light, due, for example, from an emerging supernova. We return to this possibility in Section~\ref{sec:other}.
If, instead, the late-time images are dominated by the host-galaxy light, measuring its redshift becomes critical to establishing the nature of EP250207b. The proximity to G1 suggests it could be a satellite dwarf at the same distance.
However, the observed colours ($F160W{-}F606W\!\approx\!0.6$, $F125W{-}F606W\!\approx\!0.2$) are also consistent with a higher redshift. They can be explained with a star-forming dwarf galaxy at $z\approx$1.1-1.5, which  would place the $H\alpha$ line within the $F160W$ (broad $H$) band, or with a solution at $z\gtrsim3$, which would place the Balmer/4000~\AA\ break near the $F160W$ band.

\section{Discussion} 
\label{sec:discussion}

\subsection{Constraints on kilonova emission}\label{sec:kn}

Our analysis of LXT240402A and EP250702b provides tantalizing, though not definitive, evidence for a compact-object merger origin. If the events reside at low redshift, their environments together with the stringent non-detection of any accompanying supernova strongly disfavour a massive star progenitor. By contrast, if they occurred at higher redshift, our data remain consistent with a massive star origin. 
Lacking a secure redshift measurement, the smoking gun proof
of a merger progenitor would be the identification of a kilonova component in excess of the standard afterglow emission.
At the putative redshifts of $z\simeq0.048$ and $z\simeq0.082$, an AT2017gfo-like kilonova would peak at $r\!\sim\!20.5$ and $21.7$~AB mag, respectively. 
This is readily within reach of deep ground-based imaging, provided it is not outshone by simultaneous non-thermal emission from the GRB jet.

We used the X-ray data to track the non-thermal afterglow component, whose flux evolution is empirically described as a power-law in both frequency and time, $F_{\nu} \propto t^{-\alpha} \nu^{-\beta}$. 
In the case of LXT240402A, the measured temporal slope 
$\alpha\approx$1.0 and spectral index $\beta\approx$0.7 are consistent with the closure relations for synchrotron forward shock emission \citep{Sari1998,ZhangMeszaros2004,Gao2013} and indicates that the two characteristic frequencies $\nu_m$ and $\nu_c$ lie below the optical band and above the X-ray band, respectively.
Based on this finding, we extrapolate the afterglow contribution, as characterised by X-ray data, back to optical frequencies (shaded area in Figure~\ref{fig:kn}) and find that it dominates the observed emission. 

In the case of EP250702b, a simple power-law spectrum with slope $\beta$\,$\approx$0.9 would
predict an optical flux of $\approx$22.8 AB mag at 1.2 d, in excess of the observed value of 23.3 AB \citep{Jonker2025}. 
This suggests the presence of a spectral break between optical and X-ray energies ($\nu_m$\,$\lesssim$\,$\nu_o$\,$\lesssim$\,$\nu_c$\,$\lesssim$\,$\nu_X$). 
In this regime, the temporal decay of the X-ray afterglow ($\approx$1.9) is steeper than model predictions ($\approx$1.35), which may indicate that a jet-break occurred during the time span of our observations but was not resolved due to the poorly sampled light curve. The post jet-break decay is the same below and above the cooling frequency $\nu_c$, and therefore we expect a similar fast decay at optical and nIR wavelengths. 
An early ($\lesssim$2 d) jet-break is also consistent with the lack of radio detection at later times.

Our simple modelling reveals a dominant contribution of the afterglow component, challenging the identification of any kilonova. 
In Figure~\ref{fig:kn} we compare our observational constraints with the kilonova AT2017gfo and a broad grid of kilonova models from \citet{Wollaeger_2021}. These are based on radiative transfer simulations performed with the Monte Carlo code \texttt{SuperNu} \citep{Wollaeger2013}, including the full suite of lanthanide and fourth-row element opacities.
The models adopt a two-component prescription: a low electron fraction ($Y_e$) ejecta, characteristic of tidal material expelled dynamically during merger, and a high-$Y_e$ ejecta, representing the contribution from accretion-driven winds.  
To capture the broad range of ejecta masses, we vary the mass of each component from $0.001$ to $0.1\,M_{\odot}$ \citep{Dietrich_2017, Shibata_2019}. The velocity grid spans $0.05$ to $0.3c$, consistent with the values predicted for both dynamical and wind ejecta \citep{Kasen_2017,Kawaguchi_2018}.   From the full dataset comprising 54 different viewing angles, we selected 682 spectra with 2 different viewing angles, 0 (on-axis) and 8.5 (slightly off-axis) degrees 

Figure~\ref{fig:kn} (left) presents the $r$-band light curve of LXT240402A compared to AT2017gfo, shifted to the putative redshift of $z = 0.04789$, together with the subset of models that satisfy the optical constraints. 
The optical data are fully consistent with the predictions of the afterglow model (shaded area), any kilonova contribution would thus be fainter than AT2017gfo.  These constraints disfavour models that produce bright early-time optical luminosity, primarily impacting the wind component.
The surviving models (183 out of 682, $\sim 27\%$)  
are clustered around $0.001{-}0.003\,M_\odot$ with a modal velocity of $\sim 0.3c$.
The low-$Y_e$ component remains mostly unconstrained.

Figure~\ref{fig:kn} (right) presents the $H$-band light curve of EP250207b 
compared to AT2017gfo, shifted to the putative redshift of $z = 0.082$, together with the subset of models that satisfy the near-IR constraints. 
Here, we report the $HST$ measurement derived from image subtraction (Table~\ref{tab:observationsEP250207b}), assuming that the second epoch is dominated by an underlying constant source. 
Our analysis shows that after $\gtrsim$1 d the afterglow contribution was much fainter than AT2017gfo, which would have facilitated the identification of a kilonova peak. Unfortunately, near-IR observations did not begin until 7-8 days after the burst, in part owing to the delayed announcement of this event and of its optical counterpart. 

The $F160W$ measurement at $\approx$8 d is slightly higher than our afterglow model, although within its 3 $\sigma$ range. This constraint rules out the most massive models ($\gtrsim\!0.03\,M_\odot$) and shows that any kilonova contribution would be much fainter than AT2017gfo at a similar time.  

\begin{figure}
	\includegraphics[clip, width=0.45\textwidth]{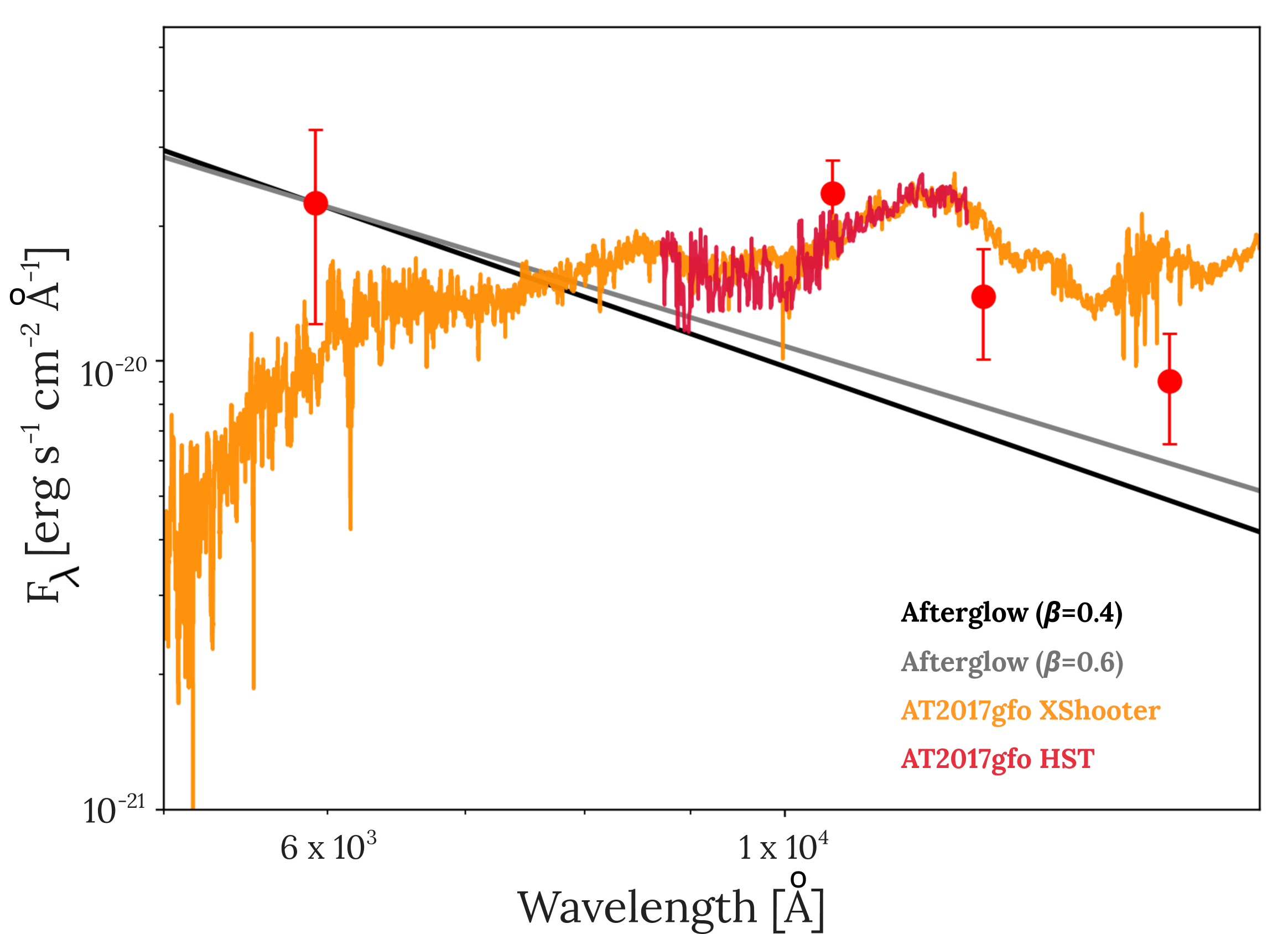}
    \caption{Optical/nIR spectral energy distribution of EP250207b at 7.4 d, compared with the X-Shooter spectrum \citep{Pian2017} and the \textit{HST} spectrum \citep{Troja2017} of AT2017gfo at a similar epoch, rescaled to match the observed photometry and redshifted to $z\sim$0.082. 
    Data were corrected for Galactic extinction using $E(B-V)\!\approx0.05$ \citep{Schlafly2011}. 
    The solid lines show the afterglow models. \label{fig:sedkn}}
\end{figure}

\subsection{An infrared excess in EP250207b}\label{sec:bump}

Our analysis demonstrates that the non-thermal afterglow radiation substantially contributes to the observed optical and nIR fluxes (Figure~\ref{fig:kn}). 
However, a simple power-law fit to the \textit{HST} data (Table~\ref{tab:observationsEP250207b}) yields  $\beta\!\approx\!0.8$ and $\chi^2\!\approx\!5$ for 2 degrees of freedom. The large residuals suggest that this model is not sufficient to describe the dataset. 
Allowing for dust along the sight line with a Milky Way–type law \citep{Cardelli1989} lowers the $\chi^2$ to $\approx$2 if $E(B{-}V)\!\approx\!0.4$, but at the same time flattens the intrinsic slope to $\beta\!\approx\!0$, inconsistent with the  afterglow model.  Moreover, this amount of reddening would be unusual in an early-type host.  
Therefore, we consider the possibility that the nIR flux is powered by an additional component, in excess of the non-thermal afterglow. 

Guided by the X-ray analysis,  we model the optical–nIR afterglow with a simple power law in time and frequencies. 
We use the temporal model (Figure~\ref{fig:kn}) to renormalise the $F160W$ measurement (Table~\ref{tab:observationsEP250207b}) at the common time of $\approx\!7.4$ d.
Then, we consider two possible spectral indices to model the afterglow contribution (Figure~\ref{fig:sedkn}): 
$\beta_{lo}\!\approx$0.4 derived from the best-fit X-ray spectral index as $\beta_{lo}\!=\!\beta_X$-0.5, and $\beta_{hi}\!\approx\!0.6$ which is consistent with a typical value of the electrons' spectral index $p\approx\!2.2$ as $\beta_{hi}\!=\!(p-1)/2$. 
No significant spectral evolution occurs in the post jet-break phase, when the cooling frequency $\nu_c \gtrsim \nu_o$ remains constant. 

Assuming a post jet-break phase, our afterglow model predicts an optical flux of $F606W$\,$\approx$\,27.8 AB mag at 7.4 d, fully consistent with our photometry. 
Therefore, we anchor our afterglow model to the optical flux at 7.4 d in order to minimise the uncertainties in the extrapolation. 
Setting a normalization of $F606W$\,$\approx\!28\pm0.5$\,AB mag and a slope of $0.4-0.6$, the nIR afterglow  lies in the range $\approx$\,28.3-27.1 AB mag, roughly a factor of two lower than the inferred flux in $F105W$ and $F125W$. 
Marginal evidence for an excess is also present in the $F160W$ band. 
The result is robust to reasonable variations of the spectral slope: adopting 
$\beta$\,$\approx$\,0.4-0.6 shifts the predicted nIR magnitudes by only 0.2–0.3 mag and does not erase evidence of a chromatic excess above the synchrotron continuum.

As shown in Figure~\ref{fig:sedkn}, this excess matches the location of the 1.1\,$\mu$m bump identified in the spectra of AT2017gfo, providing tantalizing evidence for r-process elements in the ejecta of EP250207b. These heavy elements produce an opacity window  around this wavelength where the radioactively powered emission can leak out \citep{Pognan2023}. 
However, at a similar epoch, AT2017gfo would be over an order of magnitude brighter and its spectrum slightly redder.  If the nIR excess in EP250207b is indeed kilonova powered, these properties imply a different ejecta configuration than AT2017gfo.

\subsection{Other progenitors}\label{sec:other}

Given the long duration of the X-ray prompt phase, the absence of a simultaneous short GRB, and the lack of an unambiguous kilonova signal, we broaden our progenitor search. 

Mergers of WDs with another compact-object, either an NS or a massive BH, are a plausible alternative pathway to long duration GRBs and FXRTs \citep[e.g.][]{Fryer1999,King2007,Fernandez2019,Yang2022,Lloyd2024}.
When the WD is tidally disrupted, it feeds a massive accretion disc which then launches the outflow powering the high-energy outburst \citep[e.g.][]{King2007,Margalit2016}. Host demographics should trace older stellar populations and permit sizeable offsets from galaxy light \citep{Toonen2018}, in agreement with the observations of EP250702b.

The engine timescale is set by the relatively extended WD-fed disc, so the prompt emission can display longer durations than a typical short GRB and modest isotropic energies compared to classical collapsar-driven long GRBs. 
The characteristic timescales expected for WD-BH systems, however, are orders of magnitude longer \citep{MacLeod2014,Ioka2016} than the minute-scale duration observed in these FXRTs. This discrepancy in timescales  makes WD-BH mergers an unlikely progenitor for LXT~240402A and EP250207b.

In the case of a WD-NS merger, typical ejecta masses $M_{\rm ej}\!\approx\!10^{-3}$–$10^{-1}\,M_\odot$ and velocities $v\!\approx\!0.03$–$0.2 c$ are expected, while the outflow composition ranges from Fe-group elements in wind ejecta to more neutron–rich material in tidal tails \citep{Fernandez2019}. These outflows power thermal supernova-like counterparts which are dimmer and faster evolving than Type Ia SNe, and bluer than kilonovae \citep{Fernandez2019, Zenati2020,Kaltenborn2023}. These transients may be followed by a longer-lasting ($\sim$month-long) tail of red emission \citep{Zenati2020}.

\begin{figure}
	\includegraphics[clip, width=0.45\textwidth]{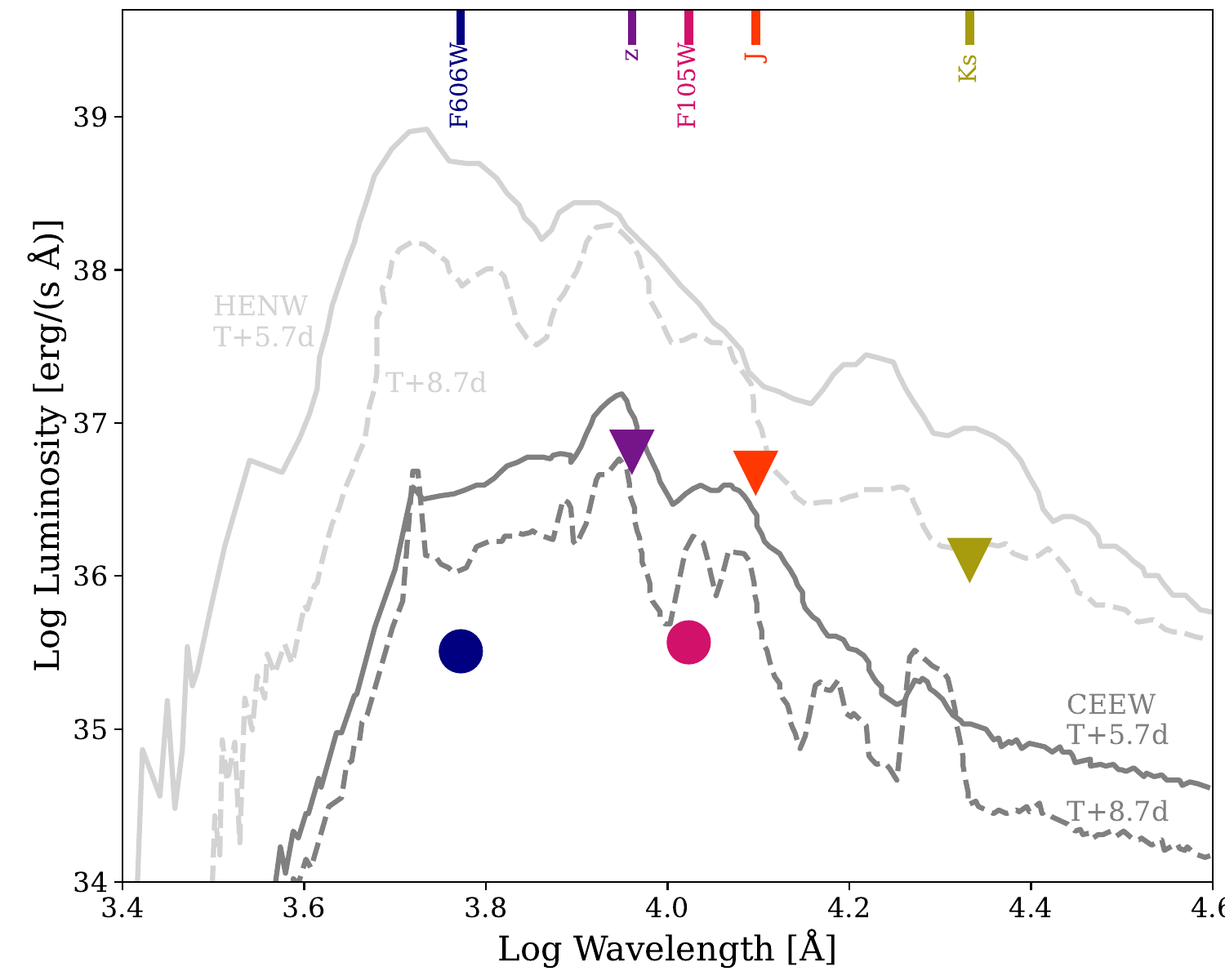} 
    \caption{Constraints on the WD-NS merger progenitor. Spectral models (high-entropy normal wind, HENW; and constant entropy normal wind, CEEW) from \citet{Kaltenborn2023} are compared to observations listed in Table~\ref{tab:observationsEP250207b}.\label{fig:hemwceew}}
\end{figure}

\begin{figure}
    \includegraphics[clip, width=0.45\textwidth]{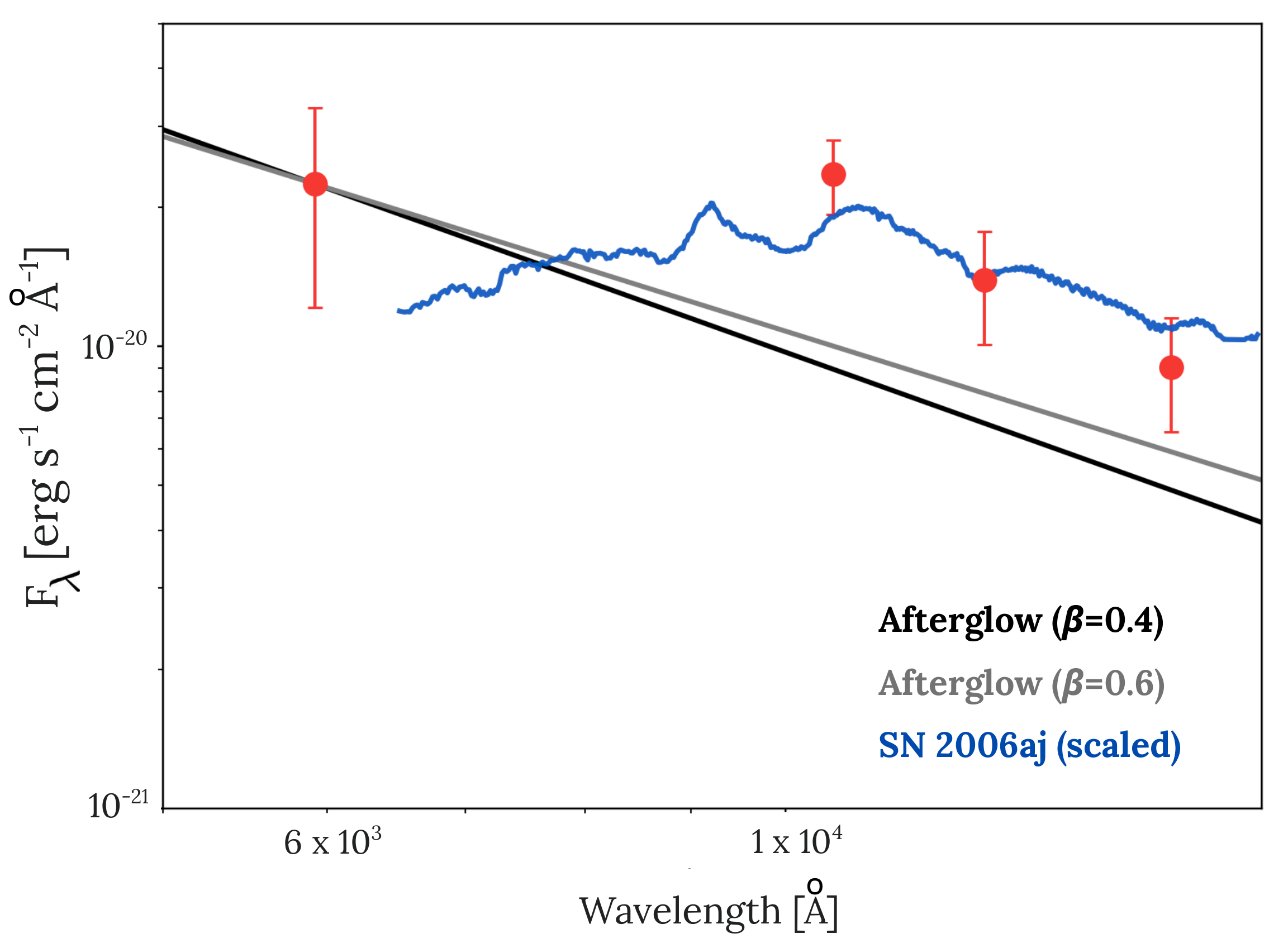} 
    \caption{Observations, same as in Fig.~\ref{fig:sedkn}, compared with the spectrum of SN2006aj at 3 days (rest-frame; \citealt{Modjaz2006}), stretched in wavelength by a factor of $1+z$ with $z$=1.15,  and arbitrarily scaled to match the observed photometry.}\label{fig:sedsn}
\end{figure}

Figure~\ref{fig:hemwceew} compares our constraints with the models of \citet{Kaltenborn2023} at 5.7 d (solid lines) and 8.7 d (dashed lines) for two distinct setup. 
The predicted transients are brighter and bluer than our data at comparable epochs and overpredict the flux in the optical bands. Barring extreme assumptions (e.g. substantial host obscuration and low ejecta mass), these comparisons disfavour a WD–NS merger as the progenitor of EP250207b.

Finally, we consider the obvious alternative of massive star progenitors, which power a large fraction of the FXRT population \citep{OConnor2025}. 
The telltale signature of a stellar core-collapse is the associated SN signal which, at a redshift $z$=0.0082, is definitely ruled out by the data. A SN would be consistent with our data set only assuming a higher redshift ($z\gtrsim$1) for EP250207b. 
We caution that the HST photometry reported in our Table~\ref{tab:observationsEP250207b} was derived via image subtraction and may not be accurate if the transient's signal persists for $\approx$30 d. For example, if the $F160W$ flux at 30 d is due to the transient rather than the galaxy, it would create an artificial dip in the spectral energy distribution. 
With this caveat in mind, we compared the spectral energy distribution (SED) at 7.4 d with the spectrum of SN2006aj at 3 d \citep{Modjaz2006}, stretched assuming $z\,\approx$1.15, and rescaled to match the observed photometry (Figure~\ref{fig:sedsn}). In this scenario, the nIR bump could be explained by typical SN spectral features 
superimposed on a standard afterglow. Therefore, without a secure distance scale to the transient, no definitive conclusion on its progenitor can be drawn.

\subsection{High-resolution X-ray spectroscopy}

The cases of LXT240402A and EP250207b illustrate the difficulties in deriving the transient's redshift from host galaxy association based exclusively on spatial proximity. 
At discovery, their optical counterparts were too faint for afterglow spectroscopy, and thus distance estimates hinge on putative hosts.

In each field, two galaxies were identified as potential hosts with similar probability of chance alignment. However, selecting one galaxy over the other would lead to radically different interpretations, from a nearby compact binary merger to a distant massive star explosion. 
In the absence of decisive discriminants, such as a GW detection or the identification of a kilonova, the classification of these events remains unsettled. 

In the following, we explore the possibility of using X-ray afterglow spectroscopy to directly infer the transient's redshift and solve similar ambiguity in future optically faint events. 
The X-ray afterglow spectrum follows a simple power-law continuum with imprinted absorption features whose energies shift according to the burst’s distance \citep{Ghisellini1999}. 
The most prominent features are the oxygen K edge 
at 0.54 keV and the iron L and K edges at 0.7-0.85 keV and 7.11 keV, respectively. 

Absorption features are visible in the spectra of GRB afterglows observed by \textit{XMM-Newton} with the Reflection Grating Spectrometers (RGS). However, only the brightest events produce marginally detectable features \citep{Campana2016}. 
We then assess the capability of \textit{XRISM}/Resolve \citep{Tashiro2025,Ishisaki2025} to identify such absorption features. Owing to its low–energy threshold of $\simeq 2~\mathrm{keV}$, the O\,K and Fe\,L edges lie below the accessible band pass, leaving only the Fe\,K edge observable for sources at $z\lesssim 2$. 

We carried out a \textsc{sixte} simulation \citep{DauserSIXTE} tailored to LXT240402A, adopting $z=1.6$ and a photon index $\Gamma=1.7$. For a flux of $5\times10^{-12}\ \mathrm{erg\ cm^{-2}\ s^{-1}}$ and a column density of $N_{\rm H}=5\times\mathrm{10}^{23}~\mathrm{cm^{-2}}$, the Fe\,K absorption edge is recovered with sufficient significance in a 100~ks exposure.
Given the observed flux of LXT240402A ($F_X$$\approx$8$\times$10$^{-13}\ \mathrm{erg\ cm^{-2}\ s^{-1}}$ at $t\approx1.5$~d), achieving successful spectroscopic constraints would require a rapid Target of Opportunity (ToO) observation, which is not among the mission's routine capabilities. Moreover, robust iron edge detection generally demands relatively large absorbing columns (e.g. $N_{\rm H}\gtrsim10^{23}\ \mathrm{cm^{-2}}$), which are only rarely encountered in long GRBs \citep{Asquini2019}. We therefore conclude that while \textit{XMM}/RGS and \textit{XRISM}/Resolve could detect the brightest FXRTs, they are unlikely to deliver spectroscopic measurements for the majority of extragalactic transients.

\begin{figure}
	\includegraphics[clip, width=0.95\linewidth]{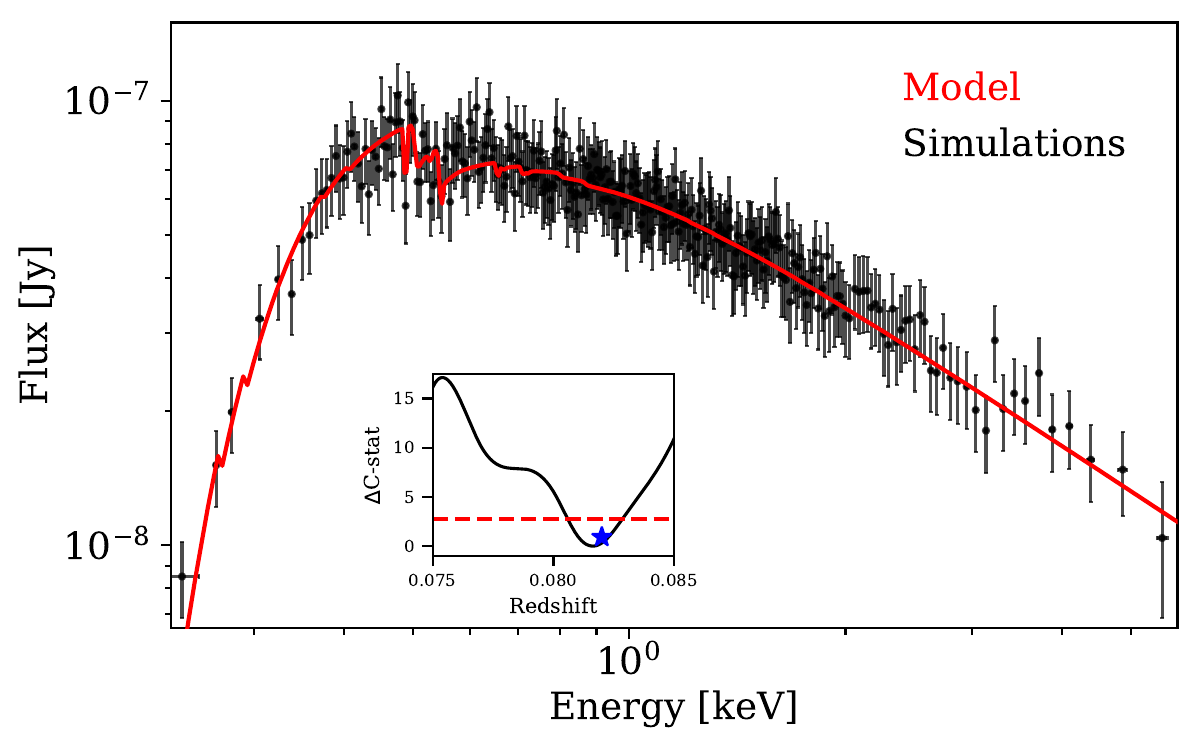}
    \caption{Simulated \textit{NewAthena} X-IFU afterglow spectrum for an integrated flux of $5 \times 10^{-13}$ erg cm$^{-2}$ s$^{-1}$ (0.3–10 keV). The best-fit absorbed power-law model is overplotted in red. The inset shows the redshift contours,  and the dashed line $\Delta$C-stat=2.706 defines the 90\% confidence level for a single free parameter \citep{Lampton76}. The star indicates the true $z$ used in the simulation. 
    \label{fig:sixte}}
\end{figure}

Looking ahead, the \textit{NewAthena} mission \citep{Cruise2025} will overcome these limitations through four key advances: a superb spectral resolution, an unprecedented collecting area, a broader energy bandpass extending to lower energies, and a rapid–response ToO program, all of which will substantially enhance the prospects for detecting spectral features in X-ray afterglows.
Figure~\ref{fig:sixte} shows a simulated X-IFU \citep{Peille2025} spectrum of EP250702b at $z$=0.082, generated with \textsc{sixte} for a 20~ks exposure and assuming a moderate intrinsic column of \(N_{\rm H}=5\times10^{20}\ \mathrm{cm^{-2}}\) and a modest X-ray flux of \(F_X=5\times10^{-13}\ \mathrm{erg\ cm^{-2}\ s^{-1}}\). 
By fitting the spectrum within XSPEC with a \texttt{tbabs $\times$ ztbabs $\times$ pegpwrlw} model, its redshift could be recovered with $\lesssim$2\% accuracy. 
This example demonstrates that the sensitivity, spectral resolution, and reaction times planned for \textit{NewAthena} would open up high-resolution X-ray spectroscopy to a broader range of transients.

\section{Summary and conclusions} 
\label{sec:summary}

We presented the case of two extragalactic FXRTs, LXT240402A and EP250702b, possibly produced by compact binary mergers.
Unlike a gamma-ray driven classification, where the transient duration defines two distinct classes of explosions \citep{Norris1984,Kouveliotou1993}, an equivalent X-ray taxonomy of extragalactic transients has yet to be established. 
The burst environment thus becomes the primary indicator of its progenitor, with massive evolved galaxies pointing to bursts from compact object mergers (also referred to as Type I bursts; \citealt{Zhang2009}) and highly star-forming galaxies generally favouring bursts from young massive stars (or Type II).
The distance scale is another key element to select high-priority targets: nearby ($z\lesssim$0.2) events allow us to robustly constrain progenitor models by identifying their hallmark signatures, such as kilonovae and supernovae. 

In this context, LXT~240402A and EP250207b represented the two most promising candidate compact binary mergers discovered by EP and its precursor LEIA. 
We obtained precise localisations of both FXRTs using the \textit{Chandra} X-ray Observatory and, on the basis of their positions onto the sky, determined that these FXRTs are possibly associated to nearby ($z\!\lesssim\!0.1$) passive galaxies, with no signs of on-going star-formation: 
LXT240402A lies in the proximity of a small galaxy group 
at $z\!\sim\!0.048$ ($P_{cc}\!\approx\!3$\%), 
EP250702b in the outskirts of a lenticular galaxy at $z\!\sim\!0.082$ ($P_{cc}\!\approx\!0.5$\%). 

However, the collected evidence in support of a compact binary merger remains circumstantial. The main sources of uncertainty are the lack of a secure distance scale and the absence of a clear kilonova signature. 
Both of these factors do not allow us to rule out a chance alignment between the nearby galaxies and the FXRTs, leaving open the possibility of a high redshift ($z\!\gtrsim\!1$) origin with a massive star progenitor \citep{OConnor2025}.  

We find that the optical emission from LXT240402A is dominated by non-thermal afterglow. 
Its sub-arcsecond localisation tends to favour a physical association with an underlying, star-forming galaxy at $z\!\approx\!1.55$. However, the probability of association with the nearby galaxy group is broadly equivalent. 
If located at $z\!\approx\!0.048$, a kilonova, slightly fainter than AT2017gfo at comparable epochs, remains consistent with the data.

The case of EP250207b is, at first glance, even more compelling. We identify a weak red excess above the standard afterglow, suggestive of a spectral feature at 1.1 $\mu$m. 
Although the spectral shape is consistent with a kilonova, its brightness is over an order of magnitude fainter than AT2017gfo at a similar epoch ($\approx$7.4 d). 
The limited dataset leaves the interpretation degenerate with a massive-star progenitor, as a supernova at higher redshift ($z\!\gtrsim\!1$) could reproduce the observations. Unfortunately, the delayed announcement of EP250207b and of its optical counterpart prevented us from securing early-time data that could have discriminated between these scenarios.

If these FXRTs were produced by compact binary mergers, our study shows that their kilonovae behave differently than AT2017gfo and other kilonovae found in short GRBs \citep{Troja2023}. 
Nonetheless, their properties remain consistent with a compact binary merger progenitor. In this framework, the energy budget of both the relativistic jet and the kilonova outflow is tied to the post-merger accretion flow \citep[e.g.][]{Gottlieb2025}. A lower accretion-disc mass naturally implies a low-luminosity jet and a fainter kilonova counterpart. At $z\!\lesssim\!0.1$ the isotropic-equivalent luminosity of LXT~240402A and EP250207b would be lower than other cosmological short GRBs, and therefore as weak kilonova component would also be expected.

We conclude by noting that X-ray afterglow spectroscopy will be especially informative for this class of optically faint transients. This capability, thus far restricted to the brightest sources, will be routinely delivered by the next-generation X-ray observatory \textit{NewAthena} \citep{Cruise2025}.

\begin{acknowledgements}
\label{sec:acknowledgments}
RB, ET, YYH, MEK, NP, MY and RR are supported by the European Research Council through the Consolidator grant BHianca (grant agreement ID~101002761). BO is supported by the McWilliams Postdoctoral Fellowship in the McWilliams Center for Cosmology and Astrophysics at Carnegie Mellon University.
AMW is grateful for support from UNAM/DGAPA through the PAPIIT project IN109224.
HS acknowledges support from the National Natural Science Foundation of China (grant No. 12573049).
AJCT acknowledges support from the Spanish Ministry project PID2023-151905OB-I00 and Junta de Andalucíia grant P20$\_$010168 and from the Severo Ochoa grant CEX2021-001131-S funded by MCIN/AEI/
10.13039/501100011033. 
MCG acknowledges financial support from the Spanish Ministry project  MCI/AEI/PID2023-149817OB-C31 and the Severo Ochoa grant  CEX2021-001131-S funded by MICIU/AEI/10.13039/501100011033. 

DS, DF, AR, AL, AT, and MU were supported by the basic funding program of the Ioffe Institute no.FFUG-2024-0002; AT acknowledges financial support from ASI-INAF Accordo Attuativo HERMES Pathfinder operazioni n. 2022-25-HH.0.

We acknowledge the support from the LBT-Italian Coordination Facility for the execution of observations, data distribution and reduction.

The scientific results reported in this article are based on observations made by the Chandra X-ray Observatory. This research has made use of software provided by the Chandra X-ray Center (CXC) in the application package CIAO. 

Based on observations collected at the European Organisation for Astronomical Research in the Southern Hemisphere under ESO programmes 114.27LW.013 and  110.24CF.021.

This work made use of data supplied by the UK \textit{Swift} Science Data Centre at the University of Leicester. 
This research has made use of the XRT Data Analysis Software (XRTDAS) developed under the responsibility of the ASI Science Data Center (ASDC), Italy. This research has made use of data and/or software provided by the High Energy Astrophysics Science Archive Research Center (HEASARC), which is a service of the Astrophysics Science Division at NASA/GSFC.

This work is based on observations made with the NASA/ESA Hubble Space Telescope. The data were obtained from the Mikulski Archive for Space Telescopes at the Space Telescope Science Institute, which is operated by the Association of Universities for Research in Astronomy, Inc., under NASA contract NAS5-03127 for JWST.

This work is based on the data obtained with Einstein Probe, a space mission supported by the Strategic Priority Program on Space Science of Chinese Academy of Sciences, in collaboration with the European Space Agency, the Max-Planck-Institute for extraterrestrial Physics (Germany), and the Centre National d'Études Spatiales (France).

This research used data obtained with the Dark Energy Spectroscopic Instrument (DESI). DESI construction and operations is managed by the Lawrence Berkeley National Laboratory. This material is based upon work supported by the U.S. Department of Energy, Office of Science, Office of High-Energy Physics, under Contract No. DE–AC02–05CH11231, and by the National Energy Research Scientific Computing Center, a DOE Office of Science User Facility under the same contract. Additional support for DESI was provided by the U.S. National Science Foundation (NSF), Division of Astronomical Sciences under Contract No. AST-0950945 to the NSF’s National Optical-Infrared Astronomy Research Laboratory; the Science and Technology Facilities Council of the United Kingdom; the Gordon and Betty Moore Foundation; the Heising-Simons Foundation; the French Alternative Energies and Atomic Energy Commission (CEA); the National Council of Humanities, Science and Technology of Mexico (CONAHCYT); the Ministry of Science and Innovation of Spain (MICINN), and by the DESI Member Institutions: www.desi.lbl.gov/collaborating-institutions. The DESI collaboration is honored to be permitted to conduct scientific research on I’oligam Du’ag (Kitt Peak), a mountain with particular significance to the Tohono O’odham Nation. Any opinions, findings, and conclusions or recommendations expressed in this material are those of the author(s) and do not necessarily reflect the views of the U.S. National Science Foundation, the U.S. Department of Energy, or any of the listed funding agencies.

This work is partly based on data obtained with the Gran Telescopio Canarias (GTC), installed in the Spanish Observatorio del Roque de los Muchachos of the Instituto de Astrofísica de Canarias, on the island of La Palma, and with the instrument OSIRIS, built by a Consortium led by the Instituto de Astrofísica de Canarias in collaboration with the Instituto de Astronomía of the Universidad Autónoma de México. OSIRIS was funded by GRANTECAN and the National Plan of Astronomy and Astrophysics of the Spanish Government. 

The National Radio Astronomy Observatory is a facility of the National Science Foundation operated under cooperative agreement by Associated Universities, Inc. The Australia Telescope Compact Array is part of the Australia Telescope National Facility which is funded by the Australian Government for operation as a National Facility managed by CSIRO. We acknowledge the Gomeroi people as the Traditional Owners of the Observatory site.
\end{acknowledgements}

%
  \bibliographystyle{aa} 
  \bibliography{references} 

\end{document}